\documentclass[preprint,showpacs,preprintnumbers,amsmath,amssymb]{revtex4}
\usepackage[abs]{overpic}
\usepackage{graphicx,axodraw,subfigure}
\begin{document}
\preprint{HUPD0802}
\def\tbr{\textcolor{red}}
\def\tcr{\textcolor{red}}
\def\ov{\overline}
\def\nn{\nonumber}
\def\f{\frac}
\def\beq{\begin{equation}}
\def\eeq{\end{equation}}
\def\bea{\begin{eqnarray}}
\def\eea{\end{eqnarray}}
\def\bsub{\begin{subequations}}
\def\esub{\end{subequations}}
\def\ynu{y_{\nu}}
\def\ydu{y_{\triangle}}
\def\ynut{{y_{\nu}}^T}
\def\ynuv{y_{\nu}\frac{v}{\sqrt{2}}}
\def\ynuvt{{\ynut}\frac{v}{\sqrt{2}}}
\def\d{\partial}
\title{CP violation of
 $\tau \to  K \pi(\eta, \eta') \nu$ decays}
\author{
Daiji Kimura$^{(a)}$,
Kang Young Lee $^{(b)}$,
T. Morozumi$^{(a)}$,
Keita Nakagawa $^{(a)}$,
}
\address{(a)
Graduate School of Science, Hiroshima University,
 Higashi-Hiroshima, 739-8526, Japan \\
(b) Department of Physics, Korea University, Seoul 136-713, Korea}
\def\nn{\nonumber}
\def\beq{\begin{equation}}
\def\eeq{\end{equation}}
\def\bei{\begin{itemize}}
\def\eei{\end{itemize}}
\def\bea{\begin{eqnarray}}
\def\eea{\end{eqnarray}}
\def\ynu{y_{\nu}}
\def\ydu{y_{\triangle}}
\def\ynut{{y_{\nu}}^T}
\def\ynuv{y_{\nu}\frac{v}{\sqrt{2}}}
\def\ynuvt{{\ynut}\frac{v}{\sqrt{2}}}
\def\s{\partial \hspace{-.47em}/}
\def\ad{\overleftrightarrow{\partial}}
\def\ss{s \hspace{-.47em}/}
\def\pp{p \hspace{-.47em}/}
\def\bos{\boldsymbol}
\begin{abstract}
We study direct CP violation of $\tau \to K \pi(\eta,\eta') \nu$
decays.
By studying the forward and backward asymmetry, the
interference of $L=0$ and $L=1$ amplitudes of the hadronic system
can be extracted. 
By including the scalar and vector mesons into
the chiral Lagrangian,
we compute the 
form factors which correspond to $L=0$ and $L=1$ angular momentum
state of the hadronic system. 
We include real and imaginary parts of the one loop corrections
to the self-energies of the scalar and vector mesons.
The direct CP violation
of the forward and backward asymmetry is computed using a
two Higgs doublet model in which a new CP
violating phase is introduced. We show how the CP
violation of the forward and 
the backward  asymmetry may
depend on the new CP violating phase and the strong phase
of the form factors.
\end{abstract}
\maketitle
\section{Introduction}
Two B factories, both Belle and Babar have accumulated 
the very large samples of $\tau$ decays.  The $\tau$
lepton physics beyond the standard model, such as
$\tau$ lepton number violation
and time reversal violation
through the measurement of electric dipole moment
have been studied. The CP violation
of the hadronic $\tau$ decay also has been investigated both
theoretically \cite{Kuhn:1996dv} and experimentally 
\cite{Bonvicini:2001xz}. Various angular 
distributions including the quantities using 
the $\tau$ spin polarization 
have been also proposed \cite{Kuhn:1996dv,Choi:1998yx}.
Recently, Belle and Babar reported the precise measurements of
the branching fractions of $ \tau \to K_s \pi \nu $ 
\cite{Belle:2007rf} and $\tau^- \to K^- \pi^0 \nu$ \cite{Aubert:2007jh}.
The improved measurement of the  
branching fraction for $ \tau \to K \eta \nu$ has been also obtained
\cite{Belle:2007mw}.
About the $\tau \to K \pi \nu $ decays, the
hadronic invariant mass spectrum has been measured.

Motivated by these measurements, we study the direct CP violation of 
$ \tau^{\pm} \to K^{\pm} P \bar{\nu} (\nu) $ with
$P=\pi^0,\eta,\eta'$ and $ \tau^{\pm} \to K_s \pi^{\pm} \bar{\nu} (\nu)$
.
Non-vanishing direct CP violation in the processes may arise 
with 
some new source of the CP violation in addition 
to Kobayashi Maskawa phase \cite{Kobayashi:1973fv}
and the strong 
phase shifts of the final states of hadrons.

In $\tau  \to K P \nu$ decays, the hadronic system
$K P$ may have the angular momentum $L=0$(s wave) and
$L=1$(p wave). The interference term of them
can be extracted from the forward and backward asymmetry 
\cite{Beldjoudi:1994hi}. In the present paper, we define
the asymmetry as the difference of the
numbers of events for K meson scattered
into the forward and backward directions
with respect to the incoming $\tau$ momentum in the
hadronic CM frame.  By comparing the forward and backward
asymmetries for CP conjugate processes,
the direct CP violation can be defined.
 The s wave and p wave of the hadronic amplitudes
are related to the scalar and vector form factors
in the time like region which have their own strong phases.
To evaluate them, we use a chiral Lagrangian including 
 scalar and  vector meson
resonances such as $\kappa(800)$ and $K^{\ast}(890)$.
We compute the  both real and imaginary parts of the one
loop corrections to the self-energy of the
resonances and obtain the 
strong phase shifts.
 We include the pseudoscalar meson loop 
correction and 
scalar and vector meson loop correction. 
The latter  may give an 
important contribution to the form factors at higher invariant mass regions above $1$ GeV
up to $m_{\tau} \sim 1.7$(GeV).

 As a new physics effect, we study a two Higgs doublet
model with non-minimal Yukawa couplings to the charged leptons.
The
two Higgs doublets contribute to the charged lepton mass through 
the Yukawa couplings. In the non-minimal
model, the interaction of the charged Higgs boson 
to the $\tau$ lepton family can be CP violating.
The
interaction generates the amplitude $\tau_R \to \nu_{\tau L} H^-
\to \nu_{\tau L} ({{\bar u}_L s_R})_{\rm s-wave} 
$.  The interference with the
charged current interaction due to $W$ boson
exchanged diagram
may lead to the direct CP violation which can be
measured in the forward
and backward asymmetry.

The paper is organized as follows. 
In section II, we show the hadronic chiral Lagrangian
including scalar and vector resonances.
In section III, we derive the form factors.
In section IV, by fixing the finite renormalization
constants, we  numerically evaluate the form factors and
the hadroic invariant mass spectrum.
In section V, we introduce
the two Higgs doublet model and
present the direct CP violation.
Section VI is devoted to conclusion and discussion.
\section{Chiral Lagrangian including scalar and vector
mesons}
In this section, we show the chiral Lagrangian with vector and
scalar resonances. The following aspects are the main feature 
of the chiral Lagrangian.
\begin{itemize}
\item{U(1)$_A$ breaking effect is taken into account so that
we can apply the Lagrangian to $\tau$ decays into
the final states including $K \eta$ and $K \eta'$.}
\item{SU(3) breaking of the vector mesons are taken into account.} 
\end{itemize}
About the inclusion of the scalar resonances, we followed 
the approach of Ref.\cite{Ecker:1988te}. About the vector meson
sector, 
our Lagrangian is equivalent to the one 
in Ref.\cite{Bando:1984ej}  except SU(3) breaking effect
for vector mesons.
The chiral Lagrangian is given by,
\bea
{\cal L}&=&\frac{f^2}{4} {\rm Tr} D U D U^{\dagger}+B 
{\rm Tr}{M (U+U^{\dagger})}
- i g_{2p} {\rm Tr} \left(\xi M \xi - \xi^{\dagger} M \xi^{\dagger}
\right)
\eta_0 -\frac{M_{0}^2}{2} \eta_0^2 \nn \\
&+& {\rm Tr}D_{\mu} S D^{\mu} S-M_{\sigma}^2 {\rm Tr}S^2 \nn \\
&+& \frac{g_1}{4}{\rm Tr}(D_{\mu} U D^{\mu} U^{\dagger})
(\xi S \xi^{\dagger})
+g_2 {\rm Tr} \left((\xi M \xi + \xi^{\dagger} M \xi^{\dagger})
S \right) \nn \\
&-&\frac{1}{2} {\rm Tr} F_{\mu \nu} F^{\mu \nu}
+ M_{V}^2 {\rm Tr}(V_{\mu}-\frac{\alpha_{\mu}}{g})^2 + g_{1V}{\rm Tr} S (V_{\mu}-\frac{\alpha_{\mu}}{g})^2,
\label{Chiral} 
\eea
where $S$ and $V$ are the scalar nonets and vector nonets
respectively. (See appendix A.)  $U$ is the chiral field and is given 
as $U=\exp(2i \pi/f)=\xi^2$. $\pi$ is SU(3) octet
pseudo Nambu Goldstone boson and $\eta_0$ corresponds to
U(1)$_{\rm A}$ pseudoscalar of which mass is denoted by
$M_0$. The U(1)$_A$ symmetry is broken by the 
mass term explicitly.
The covariant derivatives for the chiral field and the scalar 
field  are given as,
\bsub
\label{cov}
\bea
D_{\mu} U&=&(\partial_{\mu} +i A_{L \mu} )U, 
\label{covU}
\eea
\bea
D_{\mu} S&=& \partial_{\mu} S+i [\alpha_{\mu}, S],
\eea
\label{covS}
\bea
\alpha_{\mu}&=&\alpha^0_{\mu}+\xi^{\dagger} 
\frac{A_{L\mu}}{2} \xi,
\eea
\bea
\alpha^0_{\mu}&=&\frac{\xi^{\dagger} \partial_{\mu} \xi +
\xi \partial_{\mu} \xi^{\dagger}}{2i},
\eea
\esub
where $A_L$ denotes the external vector field corresponds
to SU(3)$_L$. 
$M$ in Eq.(\ref{Chiral})
is the chiral breaking term for the light quarks
and is given by,
\bea
M&=&{\rm diag.}(m_u,m_d,m_s)\nn \\
&=& m_s \cdot {\rm diag.}(\Delta,\Delta_d,1).
\eea
$\Delta$ denotes $\frac{m_u}{m_s}$ and $\Delta_d=\frac{m_d}{m_s}$.
In this work, we work in the isospin limit, $m_u=m_d$.
Below we explain how we determine the parameters in the Lagrangian
of Eq.(\ref{Chiral}).
\begin{itemize}
\item{$B$, $g_1$, $g_2$ \\
In the isospin limit,
the vacuum expectation values of the scalar fields are given as,
\bea
S_{01}=S_{02}=\frac{g_2 m_u }{M_{\sigma}^2},\quad
S_{03}= \frac{g_2 m_s}{M_\sigma^2},
\label{vev}
\eea
which leads to the
SU(3) breaking of the wave function renoramalization constants 
and the decay constants of the pseudo Nambu Goldstone bosons,
\bsub
\label{decayconst}
\bea
Z_{ij}&=&1+ g_1 \frac{S_{0i}+S_{0j}}{2 f^2}, 
\label{wavefunc}
\eea
\bea
F_K&=&f \sqrt{Z_{13}}, \quad F_\pi= f \sqrt{Z_{11}}.
\label{FkFp}
\eea
\esub
The decay constant for $\eta_8$ is written as,
\bea
F_8&=&f \sqrt{\frac{Z_{11}}{3} +\frac{2 Z_{33}}{3}} \nn \\
   &=&F_\pi \sqrt{\frac{4}{3} R^2-\frac{1}{3}},
\label{Foct}
\eea
where $R$ is $\frac{F_K}{F_\pi}$.
The generalized Gell-Mann Oakes Renner relation becomes,
\bsub
\label{GellMann}
\bea
m_K^2 F_K^2&=& (m_u+m_s) \left(2 B + g_2 (S_{01}+S_{03}) \right),
\eea
\bea
m_\pi^2 F_\pi^2&=& 2 m_u \left(2 B+  2 g_2 S_{01} \right),
\eea
\esub
which can be used to express $g_2$, $g_1$ and $B$ 
with Eq.(\ref{vev}) and Eq.(\ref{decayconst})
in terms of 
the physical quantities as, 
\begin{subequations}
\label{zenbu}
\bea
g_2 m_s &=& \frac{M_{\sigma} F_K m_K }{\sqrt{1-\Delta}} 
\sqrt{\frac{1}{1+\Delta}-R^{-2}\frac{m_\pi^2}{m_K^2}
\frac{1}{2 \Delta}}, 
\label{sub1}
\eea
\bea
g_1&=& 2 \frac{M_{\sigma}}{m_K} F_K (1-R^{-2}) 
\frac{1}{\sqrt{\frac{1}{1+\Delta}-\frac{1}{2 \Delta R^2} 
\frac{m_{\pi}^2}{m_K^2}}}\frac{1}{\sqrt{1-\Delta}},
\label{sub2} 
\eea
\bea
B m_s&=&\frac{1}{4 \Delta} \frac{1+\Delta}{1-\Delta} m_\pi^2 F_\pi^2-
\frac{1}{1+ \Delta} \frac{\Delta}{1-\Delta} m_K^2 F_K^2.
\label{sub3}
\eea
\end{subequations}
}
\item{$\eta$ and $\eta'$ mesons and octet and singlet mixing
angle $\theta_{08}$ \\
$g_{2p}$ in Eq.(\ref{Chiral})
leads to the $\eta_0$ and $\eta_8$ mixing.
The mass matrix for $\eta_0$ and $\eta_8$ sector is diagonalized 
as,
\bsub
\label{octetsinglet}
\bea
{\cal L}_{08}=-\frac{1}{2}(\eta_8, \eta_0) \left( \begin{array}{cc}
M_{88}^2 & M_{08}^2 \\
M_{08}^2 & M_{00}^2 \end{array} 
\right) \left(\begin{array}{c} \eta_8 \\
                                \eta_0 
\end{array} \right)=-\frac{1}{2} 
(\eta,\eta') \left( \begin{array}{cc}
M_{\eta}^2 & 0\\
0 & M_{\eta'}^2 \end{array} 
\right) \left(\begin{array}{c} \eta \\
                                \eta'
\end{array} \right),
\label{massmatrix}
\eea
\bea
\left( \begin{array}{c}
\eta \\
\eta' \end{array} \right)=
\left( \begin{array}{cc}
\cos \theta_{08} & -\sin \theta_{08} \\
\sin \theta_{08} & \cos \theta_{08} \end{array} \right)
\left( \begin{array}{c}
\eta_8 \\
\eta_0 \end{array} \right).
\label{mixings} 
\eea
\esub
where  $M_{08}^2=\frac{4}{\sqrt{3}}\frac{m_s-m_u}{F_8}g_{2p}$
and we take the convention $M_{08}^2<0$. Beacuse
the octet mass $M_{88}$
is given by
\bea
M_{88}^2&=&\frac{1}{3 F_8^2} \left( \frac{8}{1+\Delta} M_K^2 F_K^2
-\frac{2}{\Delta} M_{\pi}^2 F_{\pi}^2+M_{\pi}^2 F_{\pi}^2 
\right),
\label{octet}
\eea 
$M_{88}^2$ can be determined by 
$F_K$, $F_{\pi}$, $M_K$,$M_{\pi}$, and 
$\Delta$.
With $M_{88}^2$ given by Eq.~(\ref{octet}),
the parameters $M_{00}^2$ and $M_{08}^2$ are also determined
by using the masses of $\eta$ and $\eta'$ as,
\bsub
\label{mass08}
\bea
M_{00}^2&=& M_{\eta}^2+M_{\eta'}^2-M_{88}^2,
\eea
\bea
M_{08}^2&=& -\sqrt{M_{00}^2 M_{88}^2-M_{\eta}^2 M_{\eta'}^2}.
\eea
\esub
Therefore, one may 
predict the $ \eta_0$ and $\eta_8$ 
mixing angle with the relation,
\bea
\theta_{08}&=&-\frac{1}{2} 
\arctan\frac{2|M_{08}^2|}{M_{00}^2-M_{88}^2}. 
\eea
The prediction of the mixing angle is rather close to
the one experimentally extracted from $J/\psi \to \gamma \eta(\eta')$
decays.(See for example \cite{Gerard:2004gx}.)
\bsub
\label{angle}
\bea
\theta_{08 \rm th}=-22.38 (\Delta=\frac{1}{24}),\quad
-21.49 (\Delta=\frac{1}{25}),
\label{angleth}  
\eea
\bea
\theta_{08 \rm exp}&=&
-\arctan \sqrt{\frac{\Gamma[J/\psi \to \gamma \eta]}
{\Gamma[J/\psi \to \gamma \eta']}}
\left(\frac{M_{J/\psi}^2-M_{\eta'}^2}{
M_{J/\psi}^2-M_{\eta}^2}\right)^{\frac{3}{2}} \nn \\
&\simeq&-(22.36^{+1.12}_{-1.21}).
\label{angleexp} 
\eea
\esub
}
\item{Vector meson mass spectrum \\
The vector meson masses are given by the following formulae.
\bea
M_{Vij}^2=M_V^2+g_{1V}\frac{S_{0i}+S_{0j}}{2}. 
\eea
With this formulae,
the U(3) nonets vector mesons masses are given by,
\bsub
\label{vector}
\bea
M_{\rho}^2&=&M_{\omega}^2=M_V^2+ g_{1V} S_{01},
\eea
\bea
M_{\phi}^2&=&M_V^2+ g_{1V} S_{03}, 
\eea
\bea
M_{K^\ast}^2&=&M_V^2+ g_{1V}\frac{S_{01}+S_{03}}{2}.
\eea
\esub
One can fix the parameter 
$g_{1V}$ as,
\bea
g_{1V}=2\frac{M_{\phi}^2-M_{K^{\ast}}^2}{\Delta S},
\eea
where $\Delta S$ is the difference of the vaccumm expectation
values in Eq.(\ref{vev}),
\bea
\Delta S=S_{03}-S_{01}=\frac{g_2 m_s (1-\Delta)}{M_\sigma^2}.
\eea
One can also derive the following relation by using Eq.(\ref{vector}),
\bea
M_{\rho}=\sqrt{2 M_{K^*}^2-M_{\phi}^2} 
\eea
The relation leads to the prediction  $M_{\rho}=743$ MeV
which is about $-4 \%$
smaller than the measured value.
In table~\ref{table1}, we summarize the numerical values
for the parameters in the chiral Lagrangian of Eq.~(\ref{Chiral}).
}
\end{itemize}
\begin{table}
\begin{center}
\begin{tabular}{|c|c|c|c|} \hline 
$M_{\kappa}=M_{\sigma}$ (MeV) & $800$  & 840 & 760  \\ \hline
$g_2 m_s$ (MeV$^3$)& $2.65 \times 10^7$ & $2.79 \times 10^7 $
&  $2.52 \times 10^7$  \\ \hline 
$g_1$(MeV)  & $215$& $225$ & $204$ \\ \hline
$B m_s$ (MeV$^4$)
& $9.24 \times 10^8$ &$9.24 \times 10^8$ &$9.24 \times 10^8$ \\ \hline
$\Delta=\frac{m_u}{m_s}$ & $\frac{1}{25}$ & $\frac{1}{25}$ &
$\frac{1}{25}$  \\ \hline 
$\Delta S=S_{03}-S_{01}$(MeV) & 39.8& 37.9& 41.9 \\ \hline
$g_{1V}$(MeV) & 12200 & 12800& 11600 \\ \hline
$ g$ & 5.90   & 5.90  & 5.90  \\ \hline  
\end{tabular}
\end{center}
\caption{The numerical values for the parameters in the chiral
Lagrangian. We use $M_K=494$(MeV), $M_\pi=135$(MeV), $F_K=113$
(MeV)
,$F_\pi=92.2$(MeV)
and $\Gamma_{K^{\ast}}=50.8$(MeV) 
as input. $g$ is determined with the width of $K^{\ast}$.}
\label{table1}
\end{table}
\section{Form factors}
The hadronic form factors relevant for the processes $
\tau^+ \to \bar{\nu} K^+ P (P=\pi^0, \eta, \eta')$
are,
\bea
\langle K^+(p_K) P(p_P)|\bar{u} \gamma_{\mu} s|0 \rangle
=F^{K^+ P}(Q^2)q^{\mu} +\left(F_s^{K^+ P}(Q^2)- 
\frac{\Delta_{K P}}{Q^2} F^{K^+ P}(Q^2)\right) Q^{\mu},  
\eea
with $Q^{\mu}=(p_K+p_P)^{\mu}$ and $\Delta_{KP}=m_K^2-m_P^2$.
The form factor denoted by $F$ is the vector form factor
and $F_s$ is the scalar form factor. 
The form factors have been computed by using the variety of the
methods,
Ref.\cite{Li:1996md, Finkemeier:1996dh,Jamin:2006tk,Moussallam:2007qc}. 
In this work, we have used the
hadronic chiral Lagrangian including the vector and the scalar resonances
in Eq.(\ref{Chiral}).  We compute the loop corrections
to the self-energy of the vector and the scalar resonances.
The real part of the self-energy is divergent and we need to
subtract the divergence. Corresponding to the subtractions,
we have added the polynomials.
Some of the coefficients of the polynomials are 
determined by the pole positions
and the residues of the propagator
for the resonances.

To compute the form factors, let us write
the V-A charged current in terms
of hadrons. By differentiating Eq.(\ref{Chiral}) with the
external vector fields $A_L$, we obtain the current as, 
\bea
\overline{q}_{j L} \gamma_{\mu} q_{i L}
&=&-i \frac{f^2}{2} (U \partial_{\mu} U^{\dagger})_{ij}
+\frac{M_{V}^2}{g} 
\left(\xi (V_{\mu}-\frac{\alpha^0_{\mu}}{g}) \xi^{\dagger}
\right)_{ij} -i \frac{g_1}{4} \{ U \partial_{\mu} U^{\dagger}, \xi S 
\xi^{\dagger}\}_{ij} \nn \\
&+&\frac{g_{1V}}{2g} \{S, \xi(V_{\mu}-\frac{\alpha^0_{\mu}}{g})
\xi^{\dagger} \}_{ij}
-i (\xi [S,\partial_{\mu}S] \xi^{\dagger})_{ij}
-(\xi [S,[S,\alpha^0_{\mu}]] \xi^{\dagger})_{ij}.
\label{current}
\eea
We first show the results of 
the form factors for $K \pi$ final state.
\bsub
\label{formafactors}
\bea
F^{K^+ \pi^0}(Q^2)&=&\frac{1}{\sqrt{2}}
\left\{ -\frac{R+R^{-1}}{2}+ \frac{(\Delta S)^2}{2 F_K F_\pi}+
\frac{M_{K^*}^2}{2 g^2 F_K F_{\pi}}
\left(1-\frac{M_{K^*}^2}{A_R}\right) \right. \nn \\
&+& \left.
\frac{\Pi_{VS}^{\rm T}}{2 g^2 F_K F_\pi}(1-\frac{2 M^2_{K^{\ast}}}{A_R})
\right\},
\label{fv}
\eea
\bea
F_s^{K^+ \pi^0}(Q^2)&=&
\frac{\Delta_{K \pi}}{Q^2} F^{K^+ \pi^0}(Q^2) \nn \\
&+& \frac{1}{2 \sqrt{2}} \left\{ R^{-1}-R+
\frac{M_{K^*}^2}{g^2 F_K F_{\pi}} \frac{M_{K^*}^2}{A_R}
\frac{\Delta_{K \pi}(B_R D_R-C_R^2)}{(A_R+Q^2 B_R)D_R -Q^2 C_R^2}
 \right. \nn \\
&+& \left.
\frac{1}{\frac{Q^2 C_R^2}{A_R+Q^2 B_R}-D_R} g_{\kappa K \pi}(Q^2)
\frac{\Delta S}{F_K F_\pi} \right. \nn \\
&+& \left. \frac{g_{\kappa K \pi}(Q^2) M_{K^{\ast}}^2+
 M_{K^{\ast}}^2 \Delta_{K \pi} \Delta S}{g F_K F_{\pi}}
\frac{C_R}{(A_R+Q^2 B_R) D_R-Q^2 C_R^2}  \right\} \nn \\
&+& \frac{1}{2 \sqrt{2} F_K F_\pi g^2}
\left( \frac{\Delta_{K \pi}}{Q^2} 
(\Pi_{\rm VS}^{\rm L}-\Pi_{\rm VS}^{\rm T})
(1-\frac{2{M_K^{\ast}}^2}{A_R})
+ \frac{2 \Delta_{K \pi}}{A_R} \Pi_{\rm VS}^{\rm L} \right).
\label{fs}
\eea
\esub
The form factors include
the contribution of the Feynman diagrams
shown in Fig.~\ref{fig:Feyn}.
\begin{figure}[tbp]
\begin{center}
\begin{tabular}{ccc}
\begin{picture}(120,90)(0,-25)
\Text(10,63)[]{$\tau^+$}
\Text(80,-10)[]{$K^+$}
\Text(105,5)[]{$\pi^0$}
\Text(80,45)[]{$W^+$}
\Text(40,30)[]{$V_{us}$}
\ArrowLine(10,60)(60,60)
\ArrowLine(60,60)(110,60)
\Photon(60,60)(60,30){3}{3}
\ArrowLine(60,30)(90,10)
\ArrowLine(70,0)(60,30)
\Text(120,63)[]{$\bar{\nu}$}
\Text(60,-20)[]{(1-a)}
\end{picture}
&
\begin{picture}(120,70)(0,-25)
\Text(10,63)[]{$\tau^+$}
\Text(82,-13)[]{$K^+$}
\Text(105,0)[]{$\pi^0$}
\Text(80,45)[]{$W^+$}
\Text(40,30)[]{$V_{us}$}
\Text(52,25)[]{$K^{\ast}$}
\Text(74,5)[]{$K^{\ast}$}
\ArrowLine(10,60)(60,60)
\ArrowLine(60,60)(110,60)
\Photon(60,60)(60,30){3}{3}
{\SetWidth{1}
\ArrowLine(60,30)(80,10)}
\ArrowLine(80,10)(100,-10)
\ArrowLine(80,-10)(80,10)
\Text(120,63)[]{$\bar{\nu}$}
\Text(60,-20)[]{(1-b)}
\end{picture}
& 
\begin{picture}(120,70)(0,-25)
\Text(10,63)[]{$\tau^+$}
\Text(80,-11)[]{$K^+$}
\Text(105,0)[]{$\pi^0$}
\Text(80,45)[]{$W^+$}
\Text(40,30)[]{$V_{us}$}
\Text(55,25)[]{$\kappa$}
\Text(75,5)[]{$\kappa$}
\ArrowLine(10,60)(60,60)
\ArrowLine(60,60)(110,60)
\Photon(60,60)(60,30){3}{3}
{\SetWidth{1}
\ArrowLine(60,30)(80,10)}
\ArrowLine(80,10)(100,-10)
\ArrowLine(80,-10)(80,10)
\Text(120,63)[]{$\bar{\nu}$}
\Text(60,-20)[]{(1-c)}
\end{picture}
\end{tabular}
\end{center}
\end{figure}
%
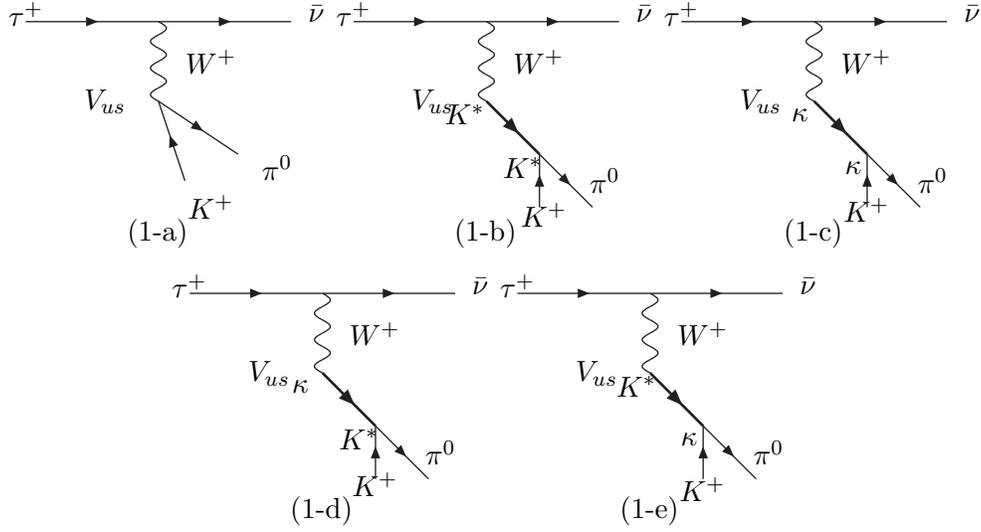
\begin{figure}[tbp]
\begin{center}
\begin{tabular}{cc}
\begin{picture}(120,70)(0,-25)
\Text(10,63)[]{$\tau^+$}
\Text(80,-12)[]{$K^+$}
\Text(105,0)[]{$\pi^0$}
\Text(80,45)[]{$W^+$}
\Text(40,30)[]{$V_{us}$}
\Text(52,25)[]{$\kappa$}
\Text(74,5)[]{$K^{\ast}$}
\ArrowLine(10,60)(60,60)
\ArrowLine(60,60)(110,60)
\Photon(60,60)(60,30){3}{3}
{\SetWidth{1}
\ArrowLine(60,30)(80,10)}
\ArrowLine(80,10)(100,-10)
\ArrowLine(80,-10)(80,10)
\Text(120,63)[]{$\bar{\nu}$}
\Text(60,-22)[]{(1-d)}
\end{picture}
& 
\begin{picture}(120,70)(0,-25)
\Text(10,63)[]{$\tau^+$}
\Text(80,-14)[]{$K^+$}
\Text(105,0)[]{$\pi^0$}
\Text(80,45)[]{$W^+$}
\Text(40,30)[]{$V_{us}$}
\Text(55,25)[]{$K^{\ast}$}
\Text(75,5)[]{$\kappa$}
\ArrowLine(10,60)(60,60)
\ArrowLine(60,60)(110,60)
\Photon(60,60)(60,30){3}{3}
{\SetWidth{1}
\ArrowLine(60,30)(80,10)}
\ArrowLine(80,10)(100,-10)
\ArrowLine(80,-10)(80,10)
\Text(120,63)[]{$\bar{\nu}$}
\Text(60,-22)[]{(1-e)}
\end{picture}
\end{tabular}
\end{center}
\caption{\label{fig:Feyn}The Feynman diagrams 
contributing to the form factors}
\end{figure}
In Fig.~\ref{fig:Feyn}, the propagators for $K^*$ and $\kappa$ mesons are
represented by the thick solid lines which  
include the one loop corrections to the self-energy.
Let us consider the 
propagators for $K^*$ and $\kappa$. They are obtained
by inverting the inverse propagators for $K^*$ and $\kappa$.
\bea
\left(\begin{array}{cc} 
(g^{\mu \nu} A_R(Q^2) + Q^{\mu} Q^{\nu} B_R(Q^2)) & Q^{\mu} C_R(Q^2) \\
Q^{\nu} C_R(Q^2) & D_R(Q^2)  \\
\end{array}
\right) \left(\begin{array}{c} 
K_{\nu} \\ 
\kappa \end{array} \right)=
\left(\begin{array}{c} J^{\mu} \\
                     J \end{array} \right),
\label{prop}
\eea
where $J_{\mu}$ and $J$ are source terms for $K^\ast$ and $\kappa$
respectively. $A_R g^{\mu \nu}+ B_R Q^{\mu} Q^{\nu}$ is the
inverse propagator for $K^{\ast}$ and $D_R$ denotes the inverse
propagator for the $\kappa$. $C_R$ denotes the mixing between
the $K^{\ast}$ and $\kappa$.
Inverting Eq.~(\ref{prop}), one can obtain the propagator.
\bea
\left(\begin{array}{c} 
K^{\mu} \\ 
\kappa 
\end{array} \right)=
\left(
\begin{array}{cc} 
g^{\mu \nu} A_R^{-1} + \frac{Q^{\mu} Q^{\nu}}{Q^2}
(\frac{D_R}{(A_R+Q^2 B_R) D_R-Q^2 C_R^2}-A_R^{-1}) & -
\frac{Q^{\mu} C_R}{(A_R+Q^2 B_R)
D_R -Q^2 C_R^2} \\
-\frac{Q^{\nu} C_R}{(A_R+Q^2 B_R)D_R -Q^2 C_R^2}  & 
\frac{A_R+Q^2 B_R}{(A_R+Q^2 B_R) D_R-Q^2 C_R^2}
\end{array} \right)
\left(\begin{array}{c} J_{\nu} \\
                     J \end{array} \right). \nn \\
\label{inverse}
\eea
To obtain the contributions to the form factors 
from Feynman diagrams  Fig.(1-a) $\sim$ Fig.(1-e),
we set $i=s,j=u$ in  Eq.~(\ref{current}),
\bea
\overline{u_L} \gamma_{\mu} s_L&=&-\frac{1}{\sqrt{2}}F_K \partial_{\mu} K^-
+ \frac{M_{K^*}^2}{\sqrt{2}g} (K^{\ast -}_{\mu}-i 
g \frac{\Delta S}{M_{K^*}^2} \partial_{\mu} \kappa^-) \nn \\
&-&i \frac{1}{2 \sqrt{2}} \left(K^- \partial_{\mu} \pi^0 
(\frac{F_{\pi}}{F_K}-\frac{M_{K^*}^2+g^2(\Delta S)^2}{2 g^2 F_K F_{\pi}})
- \partial_{\mu}  K^{-} \pi^0
 (\frac{F_{K}}{F_{\pi}}-\frac{M_{K^*}^2+g^2(\Delta S)^2}{2 g^2 F_K F_{\pi}}) \right)\nn \\
&-&i \frac{\sqrt{3}}{2 \sqrt{2}} \left(K^- \partial_{\mu} \eta_8 
(\frac{F_{8}}{F_K}-\frac{M_{K^*}^2+g^2(\Delta S)^2}{2 g^2 F_K F_{8}}) 
- \partial_{\mu}  K^{-} \eta_8
 (\frac{F_{K}}{F_{8}}-\frac{M_{K^*}^2+g^2(\Delta S)^2}{2 g^2 F_K F_{8}}) \right) \nn \\
&-&i \frac{1}{2} \left(\overline{K^0} \partial_{\mu} \pi^- 
(\frac{F_{\pi}}{F_K}-\frac{M_{K^*}^2+g^2(\Delta S)^2}{2 g^2 F_K F_{\pi}})
- \partial_{\mu}\overline{K^0} \pi^-
 (\frac{F_{K}}{F_{\pi}}-\frac{M_{K^*}^2+g^2(\Delta S)^2}{2 g^2 F_K F_{\pi}}) \right)\nn \\
&+& \frac{g_{1V}}{4g}\left(
 \frac{1}{\sqrt{2}}\kappa^- (\rho_{\mu}+\omega_{\mu})+
\overline{\kappa^0} \rho^-_{\mu}+f^0 K^{\ast-}_{\mu} \right. \nn \\
&+& \left. \frac{1}{\sqrt{2}}K^{\ast-}_{\mu} (a^0+\sigma)+
\overline{K_{\mu}^{\ast 0}} a^- +\phi_{\mu} \kappa^- \right).
\label{currents}
\eea
For the diagram in Fig.(1-a), the direct coupling of
the charged current to $K \pi$
can be easily extracted from Eq.~(\ref{currents}).
For the other diagrams, the amplitudes are obtained by
multiplying the
propagators in Eq.~(\ref{inverse})
with the production amplitudes of $K^{\ast}$ and $\kappa$ 
and the amplitudes
corresponding to their decays.

The matrix elements corresponding to Fig.(1-a) $\sim$ Fig.(1-e)
are given as,
\bea
&& \langle K^+ \pi^0|\overline{u} \gamma_{\mu} s|0 \rangle
\Large|_{\rm (1-a)}
=\frac{1}{\sqrt{2}}\left\{ Q_{\mu} \frac{R^{-1}-R}{2}+q_{\mu}
\left(-\frac{R^{-1}+R}{2}+ 
\frac{M_{K^\ast}^2+g^2 (\Delta S)^2}{2 g^2 F_K F_\pi} \right) \right\}
,\nn \\
&& \langle K^+ \pi^0|\overline{u} \gamma_{\mu} s|0 \rangle
\Large|_{\rm (1-b)}
=\frac{M_{K^\ast}^4}{2 \sqrt{2} g^2 F_K F_\pi}
\left(-q_{\mu} A_R^{-1}+ Q_{\mu} \frac{\Delta_{K \pi}}{A_R}
\frac{B_R D_R -C_R^2}{(A_R+Q^2 B_R)D_R- Q^2 C_R^2} \right)
,\nn \\
&& \langle K^+ \pi^0|\overline{u} \gamma_{\mu} s|0 \rangle
\Large|_{\rm (1-c)}
=-\frac{\Delta S g_{\kappa K \pi}(Q^2)}{2 \sqrt{2} F_K F_\pi} 
\frac{1}{D_R- \frac{Q^2 C_R^2}{A_R+Q^2 B_R}} Q_{\mu}, \nn \\
&&  \langle K^+ \pi^0|\overline{u} \gamma_{\mu} s|0 \rangle
\Large|_{\rm (1-d)}
=\frac{M_{K^{\ast}}^2}
{2 \sqrt{2} g F_K F_{\pi}}
\frac{\Delta S  \Delta_{K \pi} C_R}{(A_R+Q^2 B_R) D_R-Q^2 C_R^2} Q_{\mu}, \nn \\
&&  \langle K^+ \pi^0|\overline{u} \gamma_{\mu} s|0 \rangle
\Large|_{\rm (1-e)}
=\frac{M_{K^{\ast}}^2}
{2 \sqrt{2} g F_K F_{\pi}}
\frac{g_{\kappa K \pi}(Q^2) C_R}{(A_R+Q^2 B_R) D_R-Q^2 C_R^2} Q_{\mu}.
\label{ffs}
\eea
To derive Eq.~(\ref{ffs}), we have used  
the production amplitudes for $K^{\ast}$ and $\kappa$
due to the vector current $\overline{u} \gamma_{\mu} s$,
\bea
&& \langle K^{\ast}_{\rho} |\overline{u} \gamma_{\mu} s|0 \rangle= 
g_{\rho \mu} \sqrt{2} \frac{M_{K^{\ast}}^2}{g}, \nn \\
&&\langle \kappa |\overline{u} \gamma_{\mu} s|0 \rangle=
Q_{\mu} \sqrt{2} \Delta S.
\eea
We also have used the strong interaction vertices
which are given as,
\bea
&& \langle K^+ \pi^0|{\cal L}_{VPP}|K^{\ast +}_{\sigma}
 \rangle 
=\frac{M_{K^*}^2}{4 g F_K F_\pi} q_{\sigma}, \nn \\
&& \langle K^+ \pi^0|{\cal L}_{SPP}|\kappa^+ \rangle=
\frac{1}{4 F_K F_{\pi}} g_{\kappa K \pi}(Q^2),
\eea
with
$
q=p_K-p_\pi$. $g_{\kappa K \pi}(Q^2)$ is the strong
coupling for $\kappa \to K \pi$ defined by,
\bea
g_{\kappa K \pi}(Q^2)&=&
g_1 \frac{m_K^2+m_{\pi}^2-Q^2}{2}-g_2(3 m_u+m_s)+
\Delta_{K \pi} (\Delta S).
\eea
In addition to the pseudoscalar loops, we have taken into account
the vector and scalar meson loops denoted by $\Pi_{VS}$.
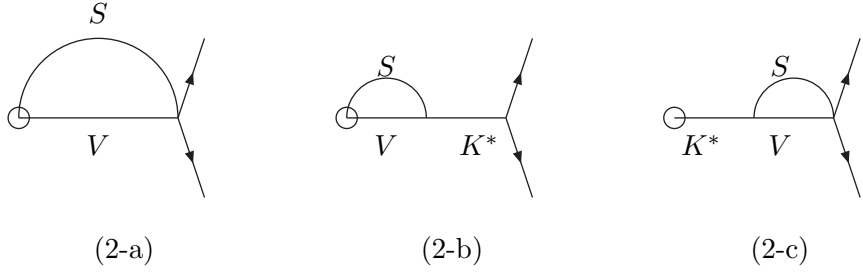
\begin{figure}[tbp]
\begin{center}
\begin{tabular}{ccc}
\begin{picture}(120,90)(0,10)
\BCirc(10,60){4}
\Line(10,60)(70,60)
\CArc(40,60)(30,0,180)
\ArrowLine(70,60)(80,90)
\ArrowLine(70,60)(80,30)
\Text(40,100)[]{$S$}
\Text(40,50)[]{$V$}
\Text(50,10)[]{(2-a)}
\end{picture}
&
\begin{picture}(120,90)(0,10)
\BCirc(10,60){4}
\Line(10,60)(70,60)
\CArc(25,60)(15,0,180)
\ArrowLine(70,60)(80,90)
\ArrowLine(70,60)(80,30)
\Text(25,80)[]{$S$}
\Text(25,50)[]{$V$}
\Text(60,50)[]{$K^{\ast}$}
\Text(50,10)[]{(2-b)}
\end{picture}
&
\begin{picture}(120,90)(0,10)
\BCirc(10,60){4}
\Line(10,60)(70,60)
\CArc(55,60)(15,0,180)
\ArrowLine(70,60)(80,90)
\ArrowLine(70,60)(80,30)
\Text(50,80)[]{$S$}
\Text(50,50)[]{$V$}
\Text(20,50)[]{$K^{\ast}$}
\Text(50,10)[]{(2-c)}
\end{picture}
\end{tabular}
\end{center}
\caption{\label{VSloop}The Feynman diagrams of scalar and 
vector mesons loop  whcih contribute to the
form factor for $ \langle K P| \bar{u} \gamma^{\mu} s|0 \rangle$.
They can be
written in terms of the self-energy correction function
$\Pi_{VS}$.} 
\end{figure}
Each contribution is given by,
\bea
\langle K^+ \pi^0|\bar{u}\gamma^{\mu} s|0 \rangle
\Large|_{(2-a)}&=&
\frac{q_{\nu} \Pi_{VS}^{\nu \mu}}{2 \sqrt{2} F_K F_\pi g^2}, \nn \\
\langle K^+ \pi^0|\bar{u}\gamma^{\mu} s|0
\rangle \Large|_{(2-b)}&=&-
\frac{M_{K^\ast}^2}{A} q_{\nu} \Pi_{VS}^{\nu \rho}
(\delta_{\rho}^{\mu}-\frac{Q_{\rho} Q^{\mu}}{M_{K^\ast}^2})
\frac{1}{2 \sqrt{2} F_K F_\pi g^2}, \nn \\
\langle
K^+ \pi^0|\bar{u}\gamma^{\mu} s|0 \rangle
\Large|_{(2-c)}&=&-
\frac{M_{K^\ast}^2}{A} q^{\rho}
(g_{\rho \nu}-\frac{Q_{\rho} Q_{\nu}}{M_{K^\ast}^2})
\frac{1}{2 \sqrt{2} F_K F_\pi g^2} \Pi_{VS}^{\nu \mu},
\eea  
where
$\Pi_{VS}$ is identical to the 
self-energy function in Fig.~(\ref{selfenergy}-d),
\bea
\Pi^{\mu \nu}_{VS}=(g^{\mu \nu}-\frac{Q^{\mu} Q^{\nu}}{Q^2})\Pi^T_{VS}+ 
\frac{Q^{\mu} Q^{\nu}}{Q^2} \Pi^L_{VS}.
\label{VS}
\eea
\begin{figure}[thbp]
\begin{center}
\begin{tabular}{cccc}
\begin{picture}(120,70)(0,10)
\Text(4,60)[]{$K^{\ast}$}
\ArrowLine(10,60)(40,60)
\CArc(55,60)(15,-90,90)
\CArc(55,60)(15,90,270)
\ArrowLine(70,60)(100,60)
\Text(110,60)[]{$K^{\ast}$}
\Text(55,80)[]{$K$}
\Text(55,30)[]{$P=\pi^0, \eta, \pi^+$}
\Text(60,10)[]{(3-a)}
\end{picture}
&
\begin{picture}(120,70)(0,10)
\Text(4,60)[]{$\kappa$}
\ArrowLine(10,60)(40,60)
\CArc(55,60)(15,-90,90)
\CArc(55,60)(15,90,270)
\ArrowLine(70,60)(100,60)
\Text(110,60)[]{$\kappa$}
\Text(55,80)[]{$K$}
\Text(55,30)[]{$P=\pi^0, \eta, \pi^+$}
\Text(60,10)[]{(3-b)}
\end{picture}
&
\begin{picture}(120,70)(0,10)
\Text(4,60)[]{$\kappa$}
\ArrowLine(10,60)(40,60)
\CArc(55,60)(15,-90,90)
\CArc(55,60)(15,90,270)
\ArrowLine(70,60)(100,60)
\Text(110,60)[]{$K^{\ast}$}
\Text(55,80)[]{$K$}
\Text(55,30)[]{$P=\pi^0, \eta, \eta' \pi^+$}
\Text(60,10)[]{(3-c)} 
\end{picture}
& \\
& 
\begin{picture}(120,80)(0,10)
\Text(4,60)[]{$K^{\ast}$}
\ArrowLine(10,60)(40,60)
\CArc(55,60)(15,-90,90)
\CArc(55,60)(15,90,270)
\ArrowLine(70,60)(100,60)
\Text(110,60)[]{$K^{\ast}$}
\Text(55,82)[]{$S$}
\Text(55,33)[]{$V$}
\Text(60,10)[]{(3-d)} 
\end{picture}
&
\end{tabular}
\end{center}
\caption{(3-a): The pseudoscalar meson loop
corrections to the self-energy
for $K^{\ast}$
($\delta A_R, \delta B_R $).  (3-b): The self-energy
for $\kappa$ ($\delta D_R$)
. (3-c): The mixing term ($C_R$). (3-d): $\Pi^{VS}$.} 
\label{selfenergy}
\end{figure}
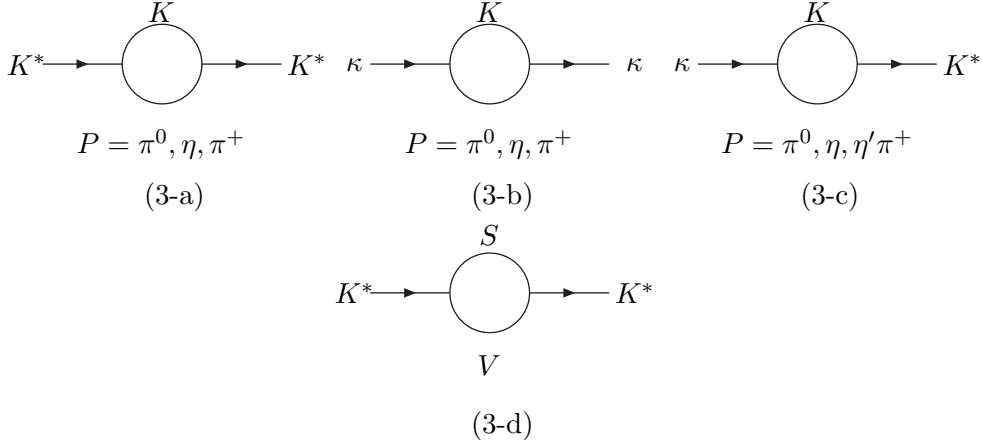
By denoting the self-energy corrections as
$\delta A_R, \delta B_R $, $C_R$ and $\Pi_{VS}$ corresponding to the
Feynman diagrams in Fig.\ref{selfenergy}, one obtains 
\bea
A_R(s)&=&M_{K^{\ast}}^2-s-i  M_{K^{\ast}}\Gamma_{K^{\ast}}(s)
+{\rm Re.} (\delta A_R(s))+ \Pi^{\rm T}_{VS}+(a_0+k_0)+(a_1+k_1)s, \nn \\
s B_R(s)&=& s(1+b_0+l_1) +i \left(\frac{\nu_{K \pi}^3}{48 \pi s^2}+
\frac{\nu_{K \pi} \Delta_{K \pi}^2}{16 \pi s^2} \right)
 3 g_{K^{\ast} K \pi}^2 \nn \\
&+& i \left(\frac{\nu_{K \eta}^3}{48 \pi s^2}+
\frac{\nu_{K \eta} \Delta_{K \eta}^2}{16 \pi s^2} \right)
 3 g_{K^{\ast} K \pi}^2 \left(\frac{F_{\pi}}{F_8}\right)^2
 \cos^2 \theta_{08}
\nn \\
&+& i \left(\frac{\nu_{K \eta'}^3}{48 \pi s^2}+
\frac{\nu_{K \eta'} \Delta_{K \eta'}^2}{16 \pi s^2} \right)
 3 g_{K^{\ast} K \pi}^2 \left(\frac{F_{\pi}}{F_8}\right)^2
 \sin^2 \theta_{08}
+ s {\rm Re.}(\delta B_R(s)) + \Pi^{\rm L}_{SV}-\Pi^{\rm T}_{SV},
\nn \\
C_R(s)&=& {\rm Re.}(C_R(s))+c_0 + i \frac{\nu_{K \pi}(s)}{16 \pi s^2}
\frac{\Delta_{K \pi}}{ 4 F_K F_{\pi}}
3 g_{\kappa K \pi}(s) 
g_{K^{\ast} K \pi} \nn \\
&-i& \frac{\nu_{K \eta}(s)}{16 \pi s^2} 
\frac{\Delta_{K \eta} F_{\pi}}{4 F_K F_8^2}
 g_{\kappa K \eta}(s) 
g_{K^{\ast} K \pi} \cos^2 \theta_{08} \nn \\
&-i& \frac{\nu_{K \eta'}(s)}{16 \pi s^2} 
\frac{\Delta_{K \eta'} F_{\pi}}{4 F_K F_8^2}
 g_{\kappa K \eta'}(s) 
g_{K^{\ast} K \pi} \sin^2 \theta_{08}, \nn \\
D_R(s)&=&s-M_{\kappa}^2 + i M_{\kappa} \Gamma_{\kappa}(s)
+{\rm Re.}(\delta D_R(s)) +d_0 + d_1 s +d_2 s^2 , 
\label{ARtoDR}
\eea
where $s=Q^2$ and 
$g_{K^{\ast} K \pi}=\frac{M_{K^{\ast}}^2}{4 g F_K F_\pi}$.
The momentum dependent
widthes $\Gamma_{K^{\ast}}(s)$ and
$\Gamma_{\kappa}(s)$ are given as,
\bea
\Gamma_{K^{\ast}}(s)&=& 3 \frac{1}{48 \pi M_{K^{\ast}}} 
\left(
\frac{\nu_{K \pi}^3}{s^2}
+ \cos^2 \theta_{08} \frac{\nu_{K \eta}^3}{s^2} 
 \left(\frac{F_{\pi}}{F_8}\right)^2 
+ \sin^2 \theta_{08} \frac{\nu_{K \eta'}^3}{s^2} 
 \left(\frac{F_{\pi}}{F_8}\right)^2
\right) g_{K^{\ast} k \pi}^2,
\nn  \\
\Gamma_{\kappa}(s)&=&3 \frac{\nu_{K \pi}}{16 \pi} 
\frac{g_{\kappa k \pi}^2(s)}{s M_{\kappa}}
\left(\frac{1}{4 F_K F_{\pi}}
 \right)^2 \nn \\
&+& \left(\cos^2 \theta_{08} \frac{\nu_{K \eta}}{16 \pi}
\frac{g_{\kappa K \eta}^2(s)}{s M_{\kappa}} 
+\sin^2 \theta_{08} \frac{\nu_{K \eta'}}{16 \pi}
\frac{g_{\kappa K \eta'}^2(s)}{s M_{\kappa}} \right)
\left(\frac{1}{ 4 \sqrt{3}  F_K F_{8}} 
 \right)^2,
\label{AtoD}
\eea
where  $\nu_{K P}=\sqrt{s^2- 2 s (m_K^2+m_P^2) +\Delta_{K P}^2}$
($P= \pi, \eta, \eta'$)
and is related to the momentum of kaon $p_K$ in the
hadronic rest frame as $p_K=\frac{\nu_{K P}}{2 \sqrt{s}}$.
The real part of the
self-energy corrections are divergent. We have subtracted
the divergences at zero momentum as,
\bea
\delta A_R(s)&=&\delta A(s)-\delta A(0) -s A'(0),\nn \\
\delta B_R(s)&=&\delta B(s)-\delta B(0), \nn \\
 C_R(s)&=& C(s)-C(0),  \nn \\
\delta D_R(s)&=&\delta D(s)-\delta D(0)-s D^{\prime}(0) -
\frac{s^2}{2} D^{\prime \prime}(0).
\label{subt}
\eea
Then we have added the polynomials with respect to $s$
which coefficients are finite renormalization constants.
We have added the polynomial $a_0 + a_1 s$ which corresponds
to the twice subtractions for $\delta A_R$ in Eq.~(\ref{subt}).
For $\delta B_R$ and $C_R$, we have added a finite
constant to each denoted by $b_0$ and $c_0$ respectively.
About the self-energy correction of the scalar meson  $
\delta D_R(s)$, we need to 
subtract divergences 
up to $s^2$. Therefore, we need to add the polynomial
$d_0 + d_1 s + d_2 s^2$ which is 
quadratic with respect to $s$.

 The self-energy corrections in Eq.~(\ref{subt}),
$\delta A_R$ $\sim $ $ C_R$, are given by,
\bea
&& \delta A_R(s)=\nn \\
&& \frac{3 g_{K^{\ast} K \pi}^2}{16 \pi^2}  \left\{
\left( -\Sigma_{K \pi}(\frac{1}{2}+\log \frac{s}{m_K m_\pi})
+\frac{s}{3}(-1+\log \frac{s}{m_K m_\pi})
- 2 s R_{K \pi}\right. \right. \nn \\ 
&-& \left. \left.
s \frac{m_K^6+m_{\pi}^6-3 m_K^2 m_\pi^2 \Sigma_{K \pi}}
{3 \Delta_{K \pi}^3} \log \frac{m_K}{m_\pi}
+
 \frac{s}{18 \Delta_{K \pi}^2}
 (5 m_K^4 +5 m_{\pi}^4-22 m_K^2 m_\pi^2)
 \right. \right. \nn \\
&+& \left. \left. 
\frac{\Sigma_{K \pi}^2+\Delta_{K \pi}^2}{4 \Delta_{K \pi}} 
\log \frac{m_K^2}{m_\pi^2}
\right) + \left(\frac{F_\pi}{F_8} \right)^2 
 \cos^2 
\theta_{08} \left( \frac{s}{18 \Delta_{K \eta}^2}
(5 m_K^4 +5 m_{\eta}^4-22 m_K^2 m_\eta^2)\right.\right. \nn \\
&-& \left. 
\left. \Sigma_{K \eta}(\frac{1}{2}+\log \frac{s}{m_K m_\eta})
+\frac{\Sigma_{K \eta}^2+\Delta_{K \eta}^2}{4 \Delta_{K \eta}} 
\log \frac{m_K^2}{m_\eta^2}
+ \frac{s}{3}(-1+\log \frac{s}{m_K m_\eta})
\right.\right. \nn \\
&-&  \left. \left. 2 s R_{K \eta}
-s 
\frac{m_K^6+m_\eta^6-3 m_K^2 m_\eta^2 
\Sigma_{K \eta}}{3 \Delta_{K \eta}^3} \log \frac{m_K}{m_\eta}
\right. \right) + 
\sin^2 \theta_{08} \left(\frac{F_\pi}{F_8} \right)^2 
\left(- 2 s R_{K \eta'}  \right. \nn \\
&-& \left. 
\Sigma_{K \eta'}(\frac{1}{2}+\log \frac{s}{m_K m_\eta'})
+\frac{s}{3}(-1+\log \frac{s}{m_K m_\eta'})
+\frac{\Sigma_{K \eta'}^2+
\Delta_{K {\eta'}}^2}{4 \Delta_{K {\eta'}}} 
\log \frac{m_K^2}{m_{\eta'}^2} \right. \nn \\
&-& \left. \left.
s \frac{m_K^6+m_\eta'^6-3 m_K^2 m_\eta'^2 
\Sigma_{K \eta'}}{3 \Delta_{K \eta'}^3}
 \log \frac{m_K}{m_{\eta'}}
+\frac{s}{18 \Delta_{K \eta'}^2} 
(5 m_K^4 +5 m_{\eta'}^4-22 m_K^2 {m_{\eta'}}^{2})
\right) \right\}, \nn \\
\eea
\bea
&& \delta B_R(s)= \nn \\
&&
\frac{3 g_{K^{\ast} K \pi}^2}{16 \pi^2}
 \left\{\left(-\frac{1}{3} \log \frac{s}{m_K m_\pi}-R'_{K \pi}
+\frac{\Sigma_{K \pi}^3}{6 \Delta_{K \pi}^3} \log \frac{m_K^2}{m_\pi^2}
-\frac{4}{9} 
\frac{m_K^4 +m_K^2 m_\pi^2+m_\pi^4}{\Delta_{K \pi}^2}
\right) +\right.
\nn \\
&& \left.
\cos^2 \theta_{08} \left(\frac{F_\pi}{F_8}\right)^2
\left(-\frac{1}{3}\log \frac{s}{m_K m_\eta}-R'_{K \eta}
+\frac{\Sigma_{K \eta}^3}{6 \Delta_{K \eta}^3} \log \frac{m_K^2}{m_\eta^2}
-\frac{4}{9} 
\frac{m_K^4 +m_K^2 m_\eta^2+m_\eta^4}{\Delta_{K \eta}^2}
\right)+  \right.\nn \\
&& \left.
\sin^2 \theta_{08} \left(\frac{F_\pi}{F_8}\right)^2
\left(-\frac{1}{3}\log \frac{s}{m_K m_{\eta'}}-R'_{K {\eta'}}
+\frac{\Sigma_{K {\eta'}}^3}{6 \Delta_{K \eta'}^3} 
\log \frac{m_K^2}{m_{\eta'}^2}
-\frac{4}{9} 
\frac{m_K^4 +m_K^2 m_{\eta'}^2+m_{\eta'}^4}{\Delta_{K {\eta'}}^2}
\right) \right\},\nn \\
\eea
\bea
&& C_R(s)= \nn \\
&&\frac{3 g_{K^\ast K \pi}}{64 \pi^2 F_K F_\pi}
\left\{-g_{\kappa K \pi}(s)(2 R^1_{K \pi}-R^0_{K \pi})
+g_{\kappa K \pi}(0)
\left( \frac{m_K^2 m_\pi^2 \log \frac{m_K^2}{m_\pi^2}}
{\Delta_{K \pi}^2}-\frac{\Sigma_{K \pi}}{2 \Delta_{K \pi}} \right)
\right\}- \nn \\
&& \cos^2 \theta_{08} \frac{3 g_{K^\ast K \pi} F_\pi}{64 \pi^2 F_K F_8^2}
\left\{-g_{\kappa K \eta}(s)(2 R^1_{K \eta}-R^0_{K \eta})
+g_{\kappa K \eta}(0) \left( \frac{m_K^2 m_\eta^2 
\log \frac{m_K^2}{m_\eta^2}}
{\Delta_{K \eta}^2}-\frac{\Sigma_{K \eta}}{2 \Delta_{K \eta}}
\right)
\right\}- \nn \\
&& \sin^2 \theta_{08} \frac{3 g_{K^\ast K \pi} F_\pi}{64 \pi^2 F_K F_8^2}
\left\{-g_{\kappa K \eta'}(s)(2 R^1_{K \eta'}-R^0_{K \eta'})
+g_{\kappa K \eta'}(0) 
\left( 
\frac{m_K^2 m_{\eta'}^2 \log \frac{m_K^2}{m_{\eta'}^2}}{\Delta_{K {\eta'}}^2}
-\frac{\Sigma_{K {\eta'}}}{2 \Delta_{K {\eta'}}} \right)
\right\},\nn \\
\eea
where $R_{PQ}$ and $R'_{PQ}$ are defined as,
\bea
R_{PQ}&=&\int_0^1 dx \left(x^2-x(1+\frac{\Delta_{PQ}}{s})+
\frac{M_P^2}{s} \right)
\log\left(x^2-x(1+\frac{\Delta_{PQ}}{s})+\frac{M_P^2}{s}-i\epsilon
\right), \nn \\
{R'}_{PQ}&=& 4 R_{PQ}^2-4 R_{PQ}^1+ R_{PQ}^0,
\eea
with
\bea
R^{(n)}_{PQ}=\int_{0}^1 dx \quad x^n
\log\left(x^2-x(1+\frac{\Delta_{PQ}}{s})+\frac{M_P^2}{s}-i\epsilon
\right).
\eea
We give the explicit forms for $R^{(n)}_{PQ}$ ($n=0 \sim 2$) 
in appendix B.
The inverse propagator for the scalar meson is,
\bea
\delta D_{R}(s)&=&\frac{3}{(4 F_K F_\pi)^2} \left\{g_{\kappa K \pi}^2(s) \bar{J}_{K \pi}(s)-g_{\kappa K \pi}^2(0) J_{K \pi}^{\prime}(0) s 
 -s^2 \left(2 g_{\kappa K \pi}(0)
g_{\kappa K \pi}^{\prime}(0) J_{K \pi}^{\prime}(0)
\right. \right. \nn \\
&+& \left. \left. \frac{1}{2}
g_{\kappa K \pi}(0)^2 J_{K \pi}^{\prime \prime}(0)
 \right) \right\} 
+\frac{\cos^2 \theta_{08}}{(4 \sqrt{3}F_K F_8)^2} 
\left\{g_{\kappa K \eta}^2(s) \bar{J}_{K \eta}(s)
-
g_{\kappa K \eta}^2(0) J_{K \eta}^{\prime}(0) s 
 \right. \nn \\
&-& \left. s^2 \left(2 g_{\kappa K \eta}(0)
g_{\kappa K \eta}^{\prime}(0) J_{K \eta}^{\prime}(0)
+ \frac{1}{2}
g_{\kappa K \eta}(0)^2 J_{K \eta}^{\prime \prime}(0) 
\right) \right\} \nn \\
&+& \frac{\sin^2 \theta_{08}}{(4 \sqrt{3}F_K F_8)^2}
\left\{g_{\kappa K \eta'}^2(s) \bar{J}_{K \eta'}(s)
-  g_{\kappa K \eta'}^2(0) J_{K \eta'}^{\prime}(0) 
s \right. \nn \\   
&-& \left. s^2 \left(2 g_{\kappa K \eta'}(0)
g_{\kappa K \eta'}^{\prime}(0) J_{K \eta'}^{\prime}(0)+\frac{1}{2}
g_{\kappa K \eta'}(0)^2
 J_{K \eta'}^{\prime \prime}(0) \right) \right\},
\eea
where 
\bea
\bar{J}_{PQ}(s)&=&
\frac{\theta(s-(M_P+M_Q)^2)}{32 \pi^2}
\left\{2+
(\frac{\Delta_{PQ}}{s} -
\frac{\Sigma_{PQ}}{\Delta_{PQ}})\log{\frac{M_Q^2}{M_P^2}}
-\frac{\nu_{PQ}}{s} \log \frac{(s+\nu_{PQ})^2-\Delta_{PQ}^2}
{(s-\nu_{PQ})^2-\Delta_{PQ}^2}  +  \right. \nn \\
&& \left.
i {2 \pi}\frac{\nu_{PQ}}{s}
\right\}+ \frac{\theta((M_P+M_Q)^2-s) 
\theta(s-(M_P-M_Q)^2)}{32 \pi^2} \left\{
(\frac{\Delta_{PQ}}{s}-
\frac{\Sigma_{PQ}}{\Delta_{PQ}}) \log{\frac{M_Q^2}{M_P^2}} \right.
\nn \\
&& \left.+2+\frac{2 \sqrt{-\nu_{PQ}^2}}{s} \left({\rm tan}^{-1} \frac{s-\Delta_{PQ}}{\sqrt{-\nu_{PQ}^2}}+{\rm tan}^{-1}
 \frac{s+\Delta_{PQ}}{\sqrt{-\nu_{PQ}^2}}\right) \right\}, 
\eea
with $\Sigma_{PQ}=m_P^2+m_Q^2$ and 
$\nu_{PQ}^2=s^2-2 s \Sigma_{PQ} +\Delta_{PQ}^2=
(s-(M_P+M_Q)^2)(s-(M_P-M_Q)^2)$. We also note,
\bea
J^{\prime}_{PQ}(0)&=& \frac{1}{32 \pi^2}
\left\{\frac{\Sigma_{PQ}}{\Delta_{PQ}^2}+2 \frac{M_P^2 M_Q^2}{\Delta_{PQ}^3}
\log \frac{M_Q^2}{M_P^2} \right\}, \nn \\
J^{\prime \prime}_{PQ}(0)&=& \frac{1}{32 \pi^2}
\left\{\frac{2}{3 \Delta_{PQ}^4}(3 \Sigma_{PQ}^2-2 \Delta_{PQ}^2)
+4 \frac{M_P^2 M_Q^2}{\Delta_{PQ}^5}
\Sigma_{PQ} \log \frac{M_Q^2}{M_P^2} \right\}.
\eea

The self-energy corrections due to
the vector and the scalar meson loop are also divergent,
\bea
{\Pi}^{\mu \nu}&=&-\frac{g_{1V}^2}{8} \sum_{VS} 
C_{VS} \mu^{4-d}
\int \frac{d^d k}{(2 \pi)^d i}
\frac{g^{\mu \nu}-\frac{k^{\mu} k^{\nu}}{m_V^2}}{(k^2-m_V^2)
((Q-k)^2-m_S^2)} 
\nn \\
&=& -\frac{g_{1V}^2}{8} \sum_{VS} C_{VS} 
(g^{\mu \nu} K_1- Q^{\mu} Q^{\nu} K_2), 
\eea
where $C_{VS}$ are factors determined by scalar and vector mesons
which contribute to the loop and are given as,
\bea
(C_{K^*\sigma},C_{\omega \kappa},
C_{\rho \kappa},
C_{K^{\ast} a},
C_{\phi \kappa},C_{K^\ast f^0})=
(\frac{1}{2},\frac{1}{2},\frac{3}{2},
\frac{3}{2},1,1).
\eea
We subtracted the divergences of $K_1$ and $K_2$ as,
\bea
K_{1 VS}&=&K_1(s)-K_1(0)-K_1^{\prime}(0) s, \nn \\
K_{2 VS}&=&K_2(s)-K_2(0).
\label{Ks}
\eea
Using $K_{1 VS}$ and $K_{2 VS}$, we write
the self-energies  
 $\Pi^T_{VS}$ and $\Pi^L_{VS}$
in Eq.~(\ref{VS}) as,
\bea
&&\Pi^T_{VS}=-\frac{g_{1V}^2}{8} \sum_{VS} C_{VS} K_{1 VS},\nn \\
&&\frac{\Pi^L_{VS}-\Pi^T_{VS}}{s}=
\frac{g_{1V}^2}{8} \sum_{VS} C_{VS} K_{2 VS}.
\eea
The explicit forms for $K_{i VS}$ $(i=1,2)$ are given as,
\bea
K_{1VS}&=&\frac{1}{16 \pi^2}\left\{
-\frac{1}{2} R^{(0)}_{VS} -\frac{s+\Delta_{VS}}{2 M_V^2} R^{(1)}_{VS}
+ \frac{s}{2 M_V^2} R^{(2)}_{VS} -1 + \frac{s}{12 M_V^2}
+\frac{\Sigma_{VS}}{8 M_V^2}- \frac{\Sigma_{VS}}{2 s} \right. \nn \\
&+& \left. \frac{s}{12 M_V^2}
\left(-\frac{3}{2}+\frac{2}{3} \frac{M_V^4 +M_V^2 M_S^2+M_S^4}{\Delta_{VS}^2}
\right)+
\log \frac{M_V M_S}{s} \left(1-\frac{\Sigma_{VS}}{4 M_V^2}+\frac{s}{12 M_V^2}
\right) \right. \nn \\\
&+&\left. \log\frac{M_V}{M_S} \left(\frac{\Sigma_{VS}}{\Delta_{VS}}
-\frac{M_V^4 +M_S^4}{4 \Delta_{VS} M_V^2}
+\frac{2 M_V^2 M_S^2}{s \Delta_{VS}} +\frac{s}{12 M_V^2}
\frac{M_V^6 + M_S^6 -3 M_V^2 M_S^2 \Sigma_{VS}}{\Delta_{VS}^3}
\right)
\right\}, \nn \\
K_{2 VS}&=& \frac{1}{16 \pi^2 M_V^2} \frac{1}{\Delta_{VS}^3}
\left\{\frac{M_V^6 \log\frac{M_V^2}{s}}{3}-\frac{11}{18}M_V^6
-\frac{M_S^6 \log\frac{M_S^2}{s}}{3}+\frac{M_S^6}{9} \right. \nn \\
&-& \left. 
\frac{M_V^2 M_S^4}{2}+M_V^4 M_S^2-M_V^2 M_S^2 \Delta_{VS}
\log\frac{M_S^2}{s} \right\}-\frac{1}{16 \pi^2 M_V^2} R^{(2)}_{VS}.
\eea
The absorptive parts of $K_{i VS}$ $(i=1,2)$ are written as,
\bea
&& {\rm Im} (K_{1VS})=\frac{\nu_{VS}}{32 \pi s}(1+\frac{s}{6 M_V^2}
+\frac{\Delta_{VS}}{2 M_V^2}+ \frac{\Sigma_{VS}}{6 M_V^2}
+\frac{\Delta_{VS}^2}{6 s M_V^2}), \nn \\
&&{\rm Im} (K_{2VS})=\frac{\nu_{VS}}{48 \pi s M_V^2}
(1-\frac{\Sigma_{VS}}{2s}+\frac{3\Delta_{VS}}{2s}+
\frac{\Delta_{VS}^2}{s^2}).
\eea
In $A_R$ and $B_R$ of Eq.(\ref{AtoD}), 
we have added the polynomial $l_1 s$ to 
 $\Pi^L_{VS}-\Pi^T_{VS}$ and $k_0+k_1 s$ to
$\Pi^T_{VS}$. The polynomials corresponds to the
subtraction of the divergent parts of Eq.~(\ref{Ks}).

The above derivation of the form factors for
$K \pi$ final state is easily extended to $K \eta$ and
$K \eta'$ case.
\bsub
\label{ffeta}
\bea
F^{K^+ \eta}(Q^2)=\cos \theta_{08} \frac{\sqrt{3}}{\sqrt{2}}
&& \left\{ -\frac{R_8+R_8^{-1}}{2}+ \frac{(\Delta S)^2}{2 F_K F_8}
+ \right. \nn \\ && \left.
\frac{\Pi_{VS}^{\rm T}}{2 g^2 F_K F_8} 
\left(1-\frac{2 M^2_{K^{\ast}}}{A_R}\right)
+
\frac{M_{K^*}^2}
{2 g^2 F_K F_{8}}
\left(1-\frac{M_{K^*}^2}{A_R}\right)
 \right\},
\label{etav}
\eea
\bea
F_s^{K^+ \eta}(Q^2)&=& 
\frac{\Delta_{K \eta}}{Q^2} F^{K^+ \eta}(Q^2)\nn \\
&+& \cos \theta_{08} \frac{\sqrt{3}}{2 \sqrt{2}} \left\{R_8^{-1}-R_8+
\frac{M_{K^*}^2}{g^2 F_K F_{8}} \frac{M_{K^*}^2}{A_R}
\frac{\Delta_{K \eta}(B_R D_R-C_R^2)}{(A_R+Q^2 B_R)D_R -Q^2 C_R^2} 
\right.\nn \\
&-& \left. 
\frac{1}{\frac{Q^2 C_R^2}{A_R+Q^2 B_R}-D_R} g_{\kappa K \eta}(Q^2)
\frac{\Delta S}{3 F_K F_{8}} \right. \nn \\
&-& \left. \frac{g_{\kappa K \eta}(Q^2) M_{K^{\ast}}^2-
 3 {M_{K^*}^2}\Delta_{K \eta}\Delta S}{3 g F_K F_{8}}
\frac{C_R}{(A_R+Q^2 B_R) D-Q^2 C^2}  \right. \nn \\
&+& \left. \frac{1}{F_K F_8 g^2}
\left(\frac{\Delta_{K \eta}}{s}(\Pi^L_{VS}-\Pi^T_{VS})
(1-\frac{2 M_{K^\ast}^2}{A_R}) +\frac{2 \Delta_{K \eta}}{A_R} \Pi^L_{VS} \right) \right\},
\label{etas}
\eea
\esub
\bsub
\label{ffetap}
\bea
F^{K^+ \eta'}(Q^2)&=&\sin \theta_{08} \frac{\sqrt{3}}{\sqrt{2}}
\left\{ -\frac{R_8+R_8^{-1}}{2}+ \frac{(\Delta S)^2}{2 F_K F_{8}}
+\frac{M_{K^*}^2}{2 g^2 F_K F_{8}}
\left(1-\frac{M_{K^*}^2}{A_R}\right)+ \right. \nn \\
&& \left.
\frac{\Pi_{VS}^{\rm T}}{2 g^2 F_K F_8}
\left(1-\frac{2 M^2_{K^{\ast}}}{A_R}\right) 
\right\},
\label{etapv} 
\eea
\bea
F_s^{K^+ \eta'}(Q^2)&=& 
\frac{\Delta_{K \eta'}}{Q^2} F^{K^+ \eta'}(Q^2)\nn \\
&+& \sin \theta_{08} \frac{\sqrt{3}}{2 \sqrt{2}} \left\{ R_8^{-1}-R_8+
\frac{M_{K^*}^2}{g^2 F_K F_{8}} \frac{M_{K^*}^2}{A_R}
\frac{\Delta_{K \eta'}(B_R D_R-C_R^2)}{(A_R+Q^2 B_R)D -Q^2 C_R^2} \right.\nn \\
&-& \left. 
 \frac{1}{\frac{Q^2 C_R^2}{A_R+Q^2 B_R}-D_R} g_{\kappa K \eta'}(Q^2)
\frac{\Delta S}
{3 F_K F_{8} } \right. \nn \\
&-& \left. \frac{g_{\kappa K \eta'}(Q^2) M_{K^{\ast}}^2-
 3 {M_{K^*}^2}\Delta_{K \eta'} \Delta S}{3  g F_K F_{8}}
\frac{C_R}{(A_R+Q^2 B_R) D_R-Q^2 C_R^2}   \right. \nn \\
&+& \left. \frac{1}{F_K F_8 g^2} \left(
\frac{\Delta_{K \eta'}}{s}(\Pi^L_{VS}-\Pi^T_{VS})
(1-\frac{2 M_{K^{\ast}}^2}{A_R}) + \frac{2 \Delta_{K \eta'}}{A_R} \Pi^L_{VS} \right) \right\},
\label{etaps}
\eea
\esub
with $R_8=\frac{F_K}{F_8}$  and 
$\Delta_{K \eta^{(')}}=m_K^2-m_{\eta^{(')}}^2$.
To derive
Eq.~(\ref{etav}) and Eq.~(\ref{etapv}), we used the strong 
interaction vertices 
for $K^{\ast} \to K \eta$ and $K^{\ast} \to K \eta'$,
\bea
\langle K^+ \eta|{\cal L}_{VPP}|K_{\sigma}^{\ast +} \rangle
&=&\cos \theta_{08} \frac{M_{K^*}^2}{4 g F_K F_8} \sqrt{3}
q_{\sigma}, \nn \\
\langle K^+ \eta'|{\cal L}_{VPP}|K_{\sigma}^{\ast +} \rangle
&=&\sin \theta_{08} \frac{M_{K^*}^2}{4 g F_K F_8} \sqrt{3}
q_{\sigma},
\eea
where $q=p_k-p_{P}$ ($P=\eta,\eta'$).
We also use
$\kappa \to K \eta^{(')}$ vertices which are 
given by,
\bea
&& \langle K^+ \eta|{\cal L}_{SPP}|\kappa^+(Q) \rangle= 
\frac{-\cos \theta_{08}}{4 \sqrt{3} F_K F_8} g_{\kappa K \eta}(Q^2), \nn \\
&& \langle K^+ \eta'|{\cal L}_{SPP}|\kappa^+(Q) \rangle=
\frac{-\sin \theta_{08}}{4 \sqrt{3} F_K F_8} g_{\kappa K \eta'}(Q^2), \nn \\
&& g_{\kappa K \eta^{(')}}(Q^2)=
g_1 \frac{m_K^2+m_{\eta^{(')}}^2-Q^2}{2}-g_2(5 m_s-m_u)- 
3 \Delta_{K \eta^{(')}} (\Delta S).
\eea
\section{Numerical analysis of the form factors}
\begin{figure}[htbp]
\begin{center}
\begin{overpic}[width=12.0cm]{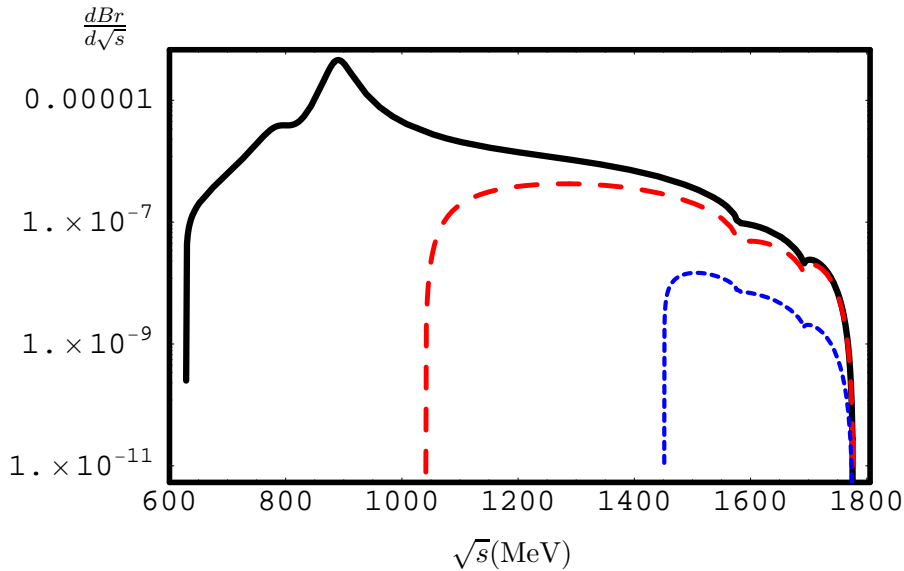}
\put(170,0){$\sqrt{s}$(MeV)}
\put(30,200){$\frac{dBr}{d\sqrt{s}}$}
\end{overpic}
\end{center}
\caption{The hadronic invariant mass
spectrum $\frac{d {\rm Br}}{d \sqrt{s}}$
for $K \pi^0$ (solid line), $K \eta$ (dashed line) and 
$K \eta'$ (short dashed line) cases.
We choose $M_{\kappa}=800$ MeV and 
the other parameters are fixed as in the corresponding
columns of Table.\ref{tableparams}.
}
\label{dBRdEH}
\end{figure}

In this section, we give the numerical results
of the form factors. We summarize the numerical values
for the masses of hadrons and width which are 
used for the numerical analysis in table~\ref{mass}.
\begin{table}
\begin{center}
\begin{tabular}{|c|c|c|c|} \hline 
$m_K$ & 493.7  & $m_{\eta^{\prime}}$& 957.8 \\ 
$m_{\pi}$& 135.0 &$M_\rho$& $775.5$ \\
$M_{K^{\ast}}$&891.7 & $M_\tau$& 1777 \\ 
$M_\phi$ & 1019 & $\Gamma_\tau$ & $2.26 \times 10^{-9}$ \\
$M_\omega$ & 782.7 & $V_{us}$ & 0.2257\\
$m_\eta$ & 547.5 &$\Gamma_{K^{\ast}}$& 50.8  \\ \hline  
\end{tabular}
\end{center}
\caption{The numerical values for masses,widths and $V_{us}$
used
in our numerical analysis. The units of mass and width are MeV.} 
\label{mass}
\end{table}
We first determine 
the finite renormalization constants. There 
are ten constants. 
We first renormalize ${\rm Re.}A_R$ and ${\rm Re.} D_R$ so that each
inverse propagator 
has
zero at the on-shell mass $s=M_{K^\ast}^2$ and $s=M_\kappa^2$
respectively.
And then we require the residues of the propagators 
$ \rm{Re.}\frac{1}{A_R}$
and ${\rm Re}\frac{1}{D_R}$ are unity on their pole masses. 
Because of the four conditions, we can constrain the 
 parameters $a_0+k_0, a_1+k_1, d_0, d_1$ and $d_2 $. 
We show the four conditions below,
\bea
&&a_0+k_0+(a_1+k_1) 
M_{K^{\ast}}^2=-{\rm Re.}\delta A_R(M_{K^{\ast}}^2)
-\rm{Re.} \Pi^T_{VS}
(M_{K^{\ast}}^2), \nn \\
&& a_1+k_1=- \frac{d {\rm Re.} (\delta A_R +\Pi^T_{VS})}{d s} 
{\Huge|}_{s={M_K^{\ast}}^2},  \nn \\
&&
d_0 +d_1 M_{\kappa}^2 +d_2 M_{\kappa}^4 
=-{{\rm Re}.\delta D_R}|_{s=M_{\kappa}^2},
\nn \\
&& 
d_1+ 2 M_{\kappa}^2
d_2 =-
\frac{d {\rm Re. \delta D_R}}{ds}|_{s=M_\kappa^2}.
\eea
Moreover we 
set four parameters $c_0,b_0,l_1,d_2$ to be zeros.
Below we show the numerical values of constants
for $M_{\kappa}=800$ MeV case.  
\bea
a_0+k_0&=&-5.68 \times  10^{5} ({\rm MeV}^2), \nn \\
a_1+k_1&=&0.345, \nn \\
d_1&=& -0.140, \nn \\
d_0&=&-1.87 \times 10^{3}({\rm MeV}^2).
\eea
We also show the values of the constants 
in Table.\ref{tableparams}
for $M_{\kappa}=760$ MeV and $840$ MeV.
Because the other parameters were set to be zero, there 
are only two undetermined parameters. We choose $k_0$
and $k_1$ as the parameters to be adjusted.  They are fixed so that
the branching fractions for $ \tau \to K \pi \nu $ and
$ \tau \to K \eta \nu $ can 
be reporoduced. For $(k_0,k_1)=(-4.03 \times 10^5,0.656)$,
we obtain 
\bea
&&Br(\tau^{\mp} \to K^{\mp} \pi \nu)=0.416 \times 10^{-2}, \nn \\
&&Br(\tau^{\mp} \to K^{\mp} \eta \nu)=1.62 \times 10^{-4},
\eea
which are close to the expeimental results 
\cite{Aubert:2007jh} and \cite{Belle:2007mw},
\bea
&& Br(\tau^{-} \to K^{-} \pi^0 \nu)=0.416 \pm 0.003\pm 0.018 \times 
10^{-2}, \nn \\
&& Br(\tau^- \to K^- \eta \nu)=(1.62\pm 0.05 \pm 0.09) \times 10^{-4}.
\eea
The branching fraction for $ \tau \to K \eta' \nu$
becomes,
\bea
{\rm Br}(\tau \to K \eta' \nu)= 3.87 \times 10^{-6}.
\eea
\begin{table}
\begin{center}
\begin{tabular}{|c|ccc|}\hline 
\begin{tabular}{|c|}
$M_\kappa$ (MeV)\\ \hline
$a_1 + k_1$ \\
$a_0 + k_0$\\
$d_1$\\
$d_0$\\
$k_0$\\
$k_1$\\ \hline 
${\rm Br}(K \pi^0)$\\ 
${\rm Br}(K \eta)$ \\ \hline
${\rm Br}(K \eta')$ 
\end{tabular}
&
\begin{tabular}{|c|}
 760\\ \hline
0.33961\\ 
 $-63519.7$\\
 $-0.139154$\\
6485.53\\
$-4.16 \times 10^5$\\
 0.6558 \\   \hline
 0.00416079\\ 
 0.000162063\\ \hline
 $4.14285 \times 10^{-6}$ 
\end{tabular}
&
\begin{tabular}{|c|}
 800 \\ \hline
0.34471 \\
$-56822.9$ \\
$-0.140467$ \\
$-1871.23$ \\
 $-4.03 \times 10^5$ \\ 
 0.6561  \\ \hline
0.00416079 \\ 
0.000162015 \\ \hline
$3.87288 \times 10^{-6}$
\end{tabular} &
\begin{tabular}{|c|} 
840 \\ \hline
0.35100 \\
$-52534.2$ \\
$-0.134301$ \\
$-16362.1$ \\
$-3.92 \times 10^5 $\\ 
 0.6564 \\ \hline
0.00415823 \\ 
0.000162325 \\ \hline
$3.94976 \times 10^{-6}$ 
\end{tabular} \\ \hline
\end{tabular} 
\end{center}
\caption{The numerical values for 
the finite renormalization constants. They
are chosen so that the the branching fractions
of $\tau \to K \pi \nu $ and $\tau \to K \eta \nu$
can be reproduced.} 
\label{tableparams}
\end{table}
We have plotted the hadronic invariant mass spectrum
for $K \pi$,$K \eta$ and $ K \eta'$ cases
in Fig.~\ref{dBRdEH}. The formulae can be found
in \cite{Kuhn:1996dv},
\bea
\frac{d {\rm Br(\tau^{\pm} \to K^{\pm} P \bar{\nu}(\nu))}}{d\sqrt{s}}
&=&\frac{p_K}{\Gamma_{\tau}} \frac{G_F^2 |V_{us}|^2}{2^5 \pi^3} 
   \frac{(m_{\tau}^2-s)^2}{m_{\tau}^3}  \nn \\
&& \left( (\frac{2 m_{\tau}^2}{3 s}+\frac{4}{3})  {p_{K}}^2  |F^{KP}(s)|^2
+\frac{m_{\tau}^2}{2} |F_s^{KP}(s)|^2 \right),
\eea
where $p_K$ is the momentum of kaon in the hadronic CM frame.
In the hadronic invariant mass spectrum for $K^{\pm} \pi^0$
at low invariant mass region, $K^{\ast}$ resonance
can be seen. Just below $K^{\ast}$, we can see
the effect of $\kappa$(800).  At the high 
invariant mass region, the new thresholds due to the 
vector and scalar channels are open and these
effects can be seen in $K \pi$, $K \eta$ and $K \eta'$
cases in Fig.~\ref{dBRdEH}.  
\section{Forward and Backward asymmetry and CP violation}
\begin{figure}[htbp]
\begin{center}
\begin{tabular}{cc}
\begin{overpic}[width=6.0cm]{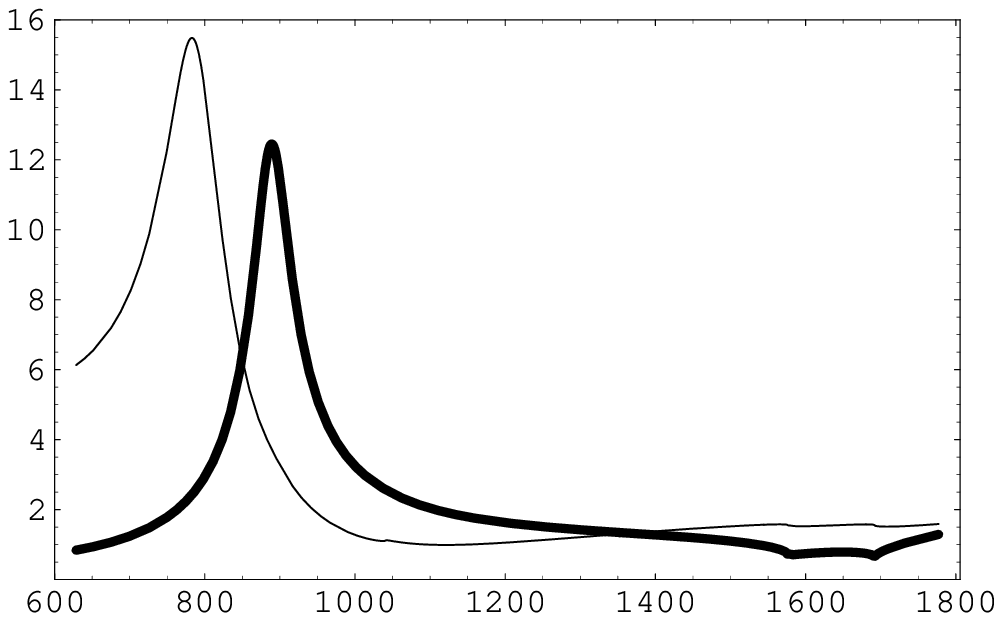}
\put(70,-10){$\sqrt{s}$(MeV)}
\end{overpic}
& \begin{overpic}[width=6cm]{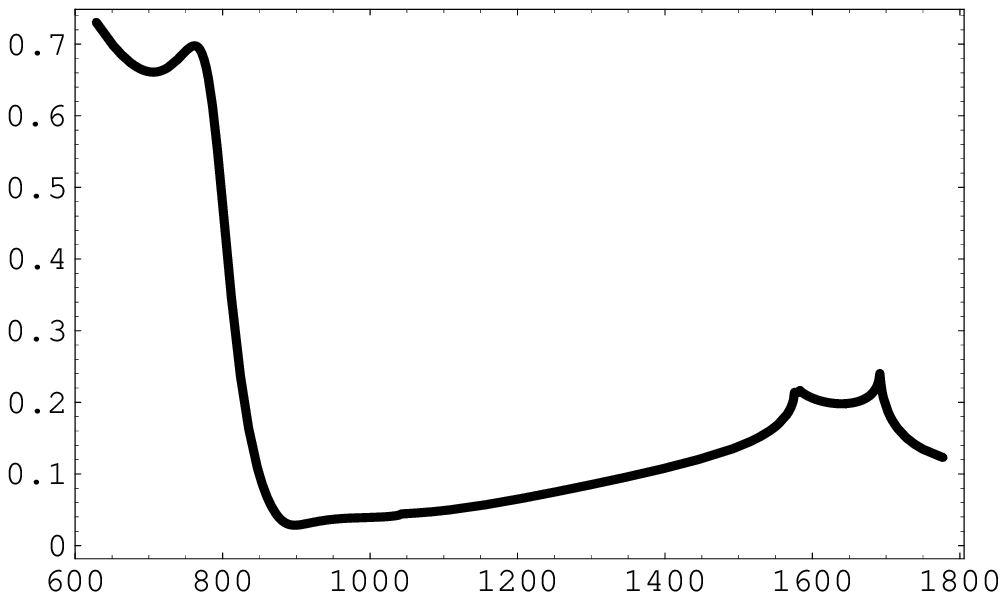}
\put(70,-10){$\sqrt{s}$(MeV)}
\end{overpic}
\end{tabular}
\end{center}
\caption{Left: The vector form factor $|F^{K \pi}|$
(thick solid line) and the scalar
form factor $ 10 \times |F_s^{K \pi}|$ (thin solid line). Right:
The ratio $\frac{|F_s^{K \pi}|}{|F^{K \pi}|}$.}
\label{Formfactors}
\end{figure}

\begin{figure}[htbp]
\begin{center}
\begin{overpic}[width=8.0cm]{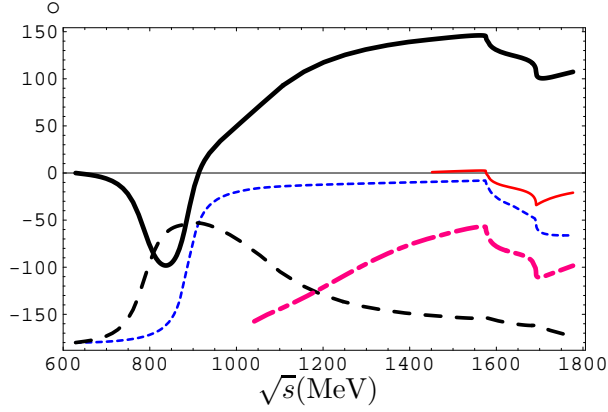}
\put(95,-5){$\sqrt{s}$(MeV)}
\put(15,140){{$\circ$}}
\end{overpic} 
\end{center}
\caption{The phase of the vector form factor: $\delta^{K \pi}_V=
{\rm arg.}F^{K \pi}$ (short dashed line) and the phase 
of the scalar form factor $\delta^{K \pi}_S={\rm arg.}F_s^{K \pi}$ 
(dashed line).
The strong phase shift
$\delta^{K \pi}_{st}=\delta^{K \pi}_V-\delta^{K \pi}_S$ is
shown with thick solid line. $\delta^{K \eta}_{st}$ and $\delta^{K \eta'}_{st}$
are shown with long short dashed line and solid line respectively. 
}
\label{phase}
\end{figure}

\begin{figure}[htbp]
\begin{center}
\begin{overpic}[width=6.0cm]{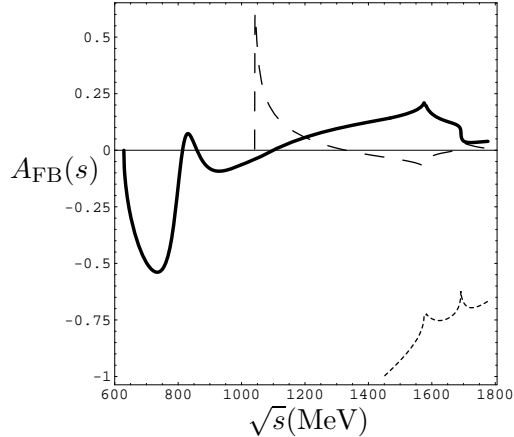}
\put(70,0){$\sqrt{s}$(MeV)}
\put(-20,95){$A_{\rm FB}(s)$}
\end{overpic}
\end{center}
\caption{The predictions 
of the forward and backward asymmtries of $\tau \to
K \pi \nu $ (thick solid line)
$\tau \to K \eta \nu $ (dashed line) and $\tau \to K \eta' \nu $ 
(short dashed line)
in the standard model.}
\label{FBASM}
\end{figure}

In this section, we compute the forward and backward
asymmetry \cite{Beldjoudi:1994hi} and the direct  CP violation
for $\tau \to K P \nu $ decay.
We start with the double differential rate of the unpolarized
$\tau$ decay \cite{Kuhn:1996dv},
\bea
\frac{d {\rm Br}}{d \sqrt{s} d \cos \theta}
&=& \frac{1}{\Gamma_{\tau}}
\frac{G_F^2|V_{us}|^2}{2^5 \pi^3} \frac{(m_{\tau}^2-s)^2}{m_{\tau}^3}
\left\{\left(\frac{m_\tau^2}{s}\cos^2 \theta
+\sin^2 \theta \right) p_K^2|F^{KP}(s)|^2+\frac{m_{\tau}^2}{4}
|F_s^{KP}|^2 \right. \nn \\
&& \left.  - \frac{m_{\tau}^2}{\sqrt{s}} p_K \cos \theta
{\rm Re.}(F^{KP}(s) F_s^{KP}(s)^{\ast}) \right\},
\eea
where $\theta$ is the scattering angle of kaon with respect to
the incoming $\tau$ in the hadronic CM frame.
The forward and backward asymmetry extracts the interference term
of the vector form factor and the scalar form factor.
\bea
A_{\rm FB}(s)&=&\frac{\int^1_0 d \cos \theta \frac{d {\rm Br}}{d \sqrt{s}
d \cos \theta}-\int^0_{-1} d \cos \theta \frac{d {\rm Br}}{d \sqrt{s}d \cos \theta}}{\frac{d {\rm Br}}{d \sqrt{s}}} \nn \\
&=&-\frac{ \frac{p_K}{\sqrt{s}}
\frac{|F_s^{KP}|}{|F^{KP}|} \cos \delta^{KP}_{\rm st} 
}{
\left(\frac{2 m_{\tau}^2}{3 s}+\frac{4}{3}\right) \frac{p_K^2}{m_{\tau}^2}
+\frac{1}{2} |\frac{F_s^{KP}}{F^{KP}}|^2 },
\label{FB}
\eea
with $\delta^{KP}_{\rm st}={\rm arg}.(\frac{F^{KP}}{F_s^{KP}}).$
As we can see from Eq.~(\ref{FB}), the forward and the backward 
asymmetry is determined by the ratio of the
scalar and the vector form factors. It is also proportional 
to cosine of the strong phase shift $\delta_{\rm st}$.
In Fig.~\ref{Formfactors}, we show the vector and the scalar
form factors and their ratio for $K \pi$ case. We also show 
the strong phase shifts in Fig.~\ref{phase}.
The forward and backward asymmetries
for $K \pi$ , $K \eta$ and $K \eta'$ cases are shown in 
Fig.~\ref{FBASM}.
As can be seen from Fig.~\ref{FBASM}, the forward
and backward asymmetry for $K \pi$ case
is large near the threshold region where 
the scalar contribution is also large. 
(See Fig.~\ref{Formfactors}.)
We can expect about $50\%$ asymmetry for $\tau \to K \pi \nu$
decay.

By including 
new physics source of CP violation
, we can predict the direct CP violation of the forward and backward asymmetry.
To be definite, we start with non-minimal two Higgs doublet model
in which the two Higgs doublets have couplings to
the charge leptons,
\bea
-{\cal L}&=&y_{1 ij} \overline{e_{Ri}} \tilde{H_1}^{\dagger}
L_{Lj}+y_{2 ij} \overline{e_{Ri}} H_2^{\dagger} L_{Lj} 
+y^d_{1 ij} \overline {d_{Ri}} \tilde{H_1}^{\dagger} Q_{Lj} \nn \\
&+&y^u_{2 ij} \overline{u_{Ri}} \tilde{H_2}^{\dagger} Q_{Lj}
+ y^{\nu}_{2ij} \overline{\nu_{Ri}} \tilde{H_2}^{\dagger} L_{Lj}
+ \frac{M_{N_i}}{2} \overline{\nu_{Ri}} (\nu_{Ri})^c
+ {\rm h.c.},
\eea
where we assume that $H_2$ is coupled with neutrinos
and both $H_1$ and $H_2$ are coupled with the charged leptons.
The neutrino mass is given by the seesaw mechanism. However,
 the right-handed Majorana neutrino with the 
mass $M_N$ much heavier than the electro-weak scale does
not affect on the interaction terms with the mass dimension
equal to four at all. Therefore, we keep the terms which are not
suppressed by a factor $\frac{1}{M_N}$ in discussion below.
In two Higgs doublet model, without loss of generality, one
can parametrize the Higgs fields as,
\bea
H_2=e^{i \frac{\theta_{CP}}{2}}\left( \begin{array}{c}
-\cos \beta H^+ \\
\frac{v_2+h_2-i \cos \beta A}{\sqrt{2}} \\
\end{array} \right), \quad 
\tilde{H_1}=i \tau_2 H_1^{\ast}
=e^{-i \frac{\theta_{CP}}{2}}\left( \begin{array}{c}
-\sin \beta H^+ \\
-\frac{v_1+h_1+i \sin \beta A}{\sqrt{2}} \\
\end{array} \right),
\eea
where $\theta_{CP}$ is the relative phase of the vaccuum expectaion
values of Higgs and its value can be determined from Higgs 
potential. The charged
current interactions of the model are,
\bea
{\cal L}&=&H^+ \overline{\nu_{L i}} l_{Rj}
(\frac{Y_{2 ji}^{\ast} e^{+i \frac{\theta}{2}}}{\cos \beta}
-\delta_{ij} \frac{g \tan \beta m_{j} } {\sqrt{2} M_W}) \nn \\
&-& H^- \overline{d_{iR}} V_{ui}^{\ast} u_L 
\frac{g m_{di}\tan \beta}{\sqrt{2} M_W}- M_H^2 H^+ H^- \nn \\
&-&\frac{g}{\sqrt{2}} V_{ui}^{\ast}
W^-_{\mu} \overline{d_{Li}} \gamma^{\mu} u_L
-\frac{g}{\sqrt{2}} 
\overline{\nu_{i L}}\gamma_{\mu} l_{Li} W^{+ \mu} + M_W^2 W^{+ \mu}
W_{- \mu}+ {\rm h.c.},
\label{chargedhiggs}
\eea
where the charged lepton masses 
$m_l={\rm diagonal}(m_e, m_{\mu}, m_{\tau})$
are obtained by the diagonalization.
\bea
V_R \frac{1}{\sqrt{2}} (-y_1 v_1 e^{i \frac{\theta_{CP}}{2}}
+y_2 v_2 e^{-i \frac{\theta_{CP}}{2}} ) V_L^{\dagger}
=m_l.
\label{chargedlepton}
\eea
Eq.~(\ref{chargedlepton}) can be used to express $y_1$
in terms of the charged lepton mass and the other
Yukawa coupling $y_2$. By introducing,
\bea
Y_2= V_R  y_2 V_L^{\dagger},
\eea
one can obtain Eq.~(\ref{chargedhiggs}).
The four fermi interactions induced by the charged Higgs exchange
are, 
\bea
{\cal L}_H&=& \frac{G_F}{\sqrt{2}} \frac{\tan^2 \beta}{M_H^2} 
 \left\{ \overline{\nu_{i}}  \left(\delta_{ij}- 
r_{2 ij}
\frac{1}{\sin \beta} \right)
m_{lj} (1+\gamma_5) l_j \right \} \nn \\
&& \times \left\{\overline{d}V_{KM}^{\dagger} \left(m_d(1-\gamma_5)+
m_u \cot^2 \beta (1+\gamma_5) \right) u \right\}+{\rm h.c.},
\eea
where $r_{2 ij}$ denotes the non-minimal couplings of 
charged Higgs boson between the charged 
lepton $l_j$ to the neutrino $\nu_i$, 
\bea
r_{2 ij}=\frac{Y_{2 ji}^{\ast} 
e^{i \frac{\theta_{CP}}{2}}}{Y_j^{SM}},
\eea
where $Y^{SM}$ denotes the standard model Yukawa couplings
for charged leptons, 
$
Y_{j}^{SM} \frac{v}{\sqrt{2}}=m_{lj}
$. The amplitude of the two Higgs doublet model is,
\bea
&&{\rm Amp.}( \tau^- \to \nu_i K^- P)
=-\frac{G_F}{\sqrt{2}} V_{us}^{\ast}
 \bar{u_i} \gamma^{\mu}
(1-\gamma_5) u_{\tau}  \nn \\
&& \left\{
\delta_{i \tau} (q_{\mu}-\frac{\Delta_{KP}}{Q^2} Q_{\mu})
F^{KP} +\left( \delta_{i \tau} 
(1-\frac{Q^2}{M_H^2} \tan^2 \beta)
+\frac{Q^2}{M_H^2} \frac{\tan^2 \beta}{\sin \beta}r_{2 i \tau}
 \right) F_s^{KP} Q_{\mu} \right\},
\eea
where the matrix element of the scalar current is given
by, 
\bea
\langle K^- \pi^0| \bar{s} u |0 \rangle= 
\frac{Q^2}{m_s-m_u} F_s^{KP}.
\eea
Then $\tau \to K P \nu_i$ branching fraction
is,
\bea
\frac{d Br(\tau^- \to K^- P \nu_i)}{d \sqrt{s} d \cos \theta}
&=&\frac{1}{\Gamma} \frac{G_F^2}{2^5 \pi^3}
\frac{(m_{\tau}^2-s)^2}{m_{\tau}^3}
\left\{ \left(\frac{m_\tau^2}{s}\cos^2 \theta
+\sin^2 \theta \right)p_K^2 |F^{KP}|^2 \delta_{i \tau} +\frac{m_{\tau}^2}{4}
|F_{s i \tau}^{KP}|^2  \right. \nn \\
&-& \left. \delta_{i \tau} \frac{m_\tau^2}{\sqrt{s}} p_K \cos \theta
{\rm Re.}(F^{KP} F_{s  \tau \tau}^{KP \ast} ) \right\},
\eea
where we define
\bea
F^{KP}_{s i \tau}=
\left\{\delta_{i \tau}\left(1-\frac{Q^2}{M_H^2} \tan^2 \beta\right)
+\frac{Q^2}{M_H^2} \tan^2 \beta \frac{r_{2 i \tau}}{\sin \beta} \right\}
F^{KP}_s
\eea
for $i=$ e, $\mu$ and $\tau$.
We neglect the small corrections proportional to 
up quark mass.
For CP conjugate processes $\tau^+ \to K^+ \pi^0 \bar{\nu_i}$
is obrained  by replacing $r_2$ in the amplitude
$\tau^- \to K^- \pi^0 \nu_i$ 
with its complex conjugate $r_2^{\ast}$. 
Therefore the direct CP violation of the forward and backward
asymmetries
is given as,
\bea
A_{FB}-\bar{A}_{FB}&=&
\frac{2 \frac{p_K}{\sqrt{s}} \frac{|F_s^{KP}|}{|F^{KP}|}
\sin \delta_{st}
}
{(\frac{2 m_{\tau}^2}{3 s}+\frac{4}{3}) \frac{p_K^2}{m_\tau^2}
+\frac{1}{2} 
\sum_{i} \frac{|F^{KP}_{s i \tau}|^2}{|F^{KP}_{\tau \tau}|^2}}
\left(\frac{Q^2}{M_H^2} \frac{\tan^2 \beta}{\sin \beta}\right)
|r_{2 \tau \tau}|\sin \theta_{2 \tau \tau},
\label{FBACPE}
\eea
where we parametrize CP violating phase of the flavor
diagonal coupling as,
\bea
r_{2 \tau \tau}=|r_{2 \tau \tau}| e^{i \theta_{2 \tau \tau}}.
\eea
We set the flavor off-diagonal couplings in $r_{2 i \tau}$
to be zeros.
We note that in the isospin limit $m_u=m_d$,
the CP asymmetry of Eq.~(\ref{FBACPE}) for $K^{\pm} \pi^0$ case 
is identical to the direct
CP violation 
of $\tau^- \to K_s \pi^- \nu_i $ and $\tau^+ \to K_s \pi^+ \bar{\nu_i}$.
Contrary to the CP violation 
of the total branching ratios which is sensitive to the
CP violation of the mixing of $K^0$ and $\bar{K^0}$
\cite{Bigi:2005ts}, 
the CP violation of the forward and backward asymmetries
does not depend on the mixing induced CP violation.
 
\begin{figure}[hbp]
\begin{center}
\begin{overpic}[width=9cm]{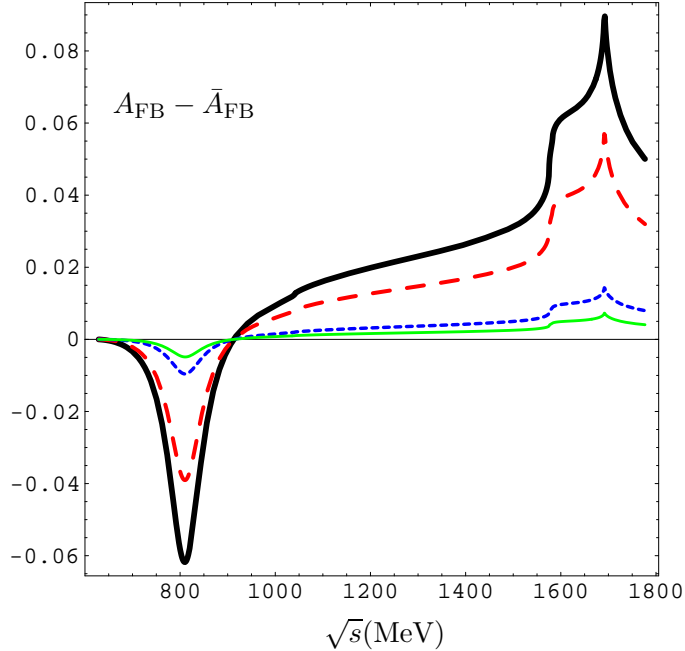}
\put(120,0){$\sqrt{s}$(MeV)}
\put(40,200){$A_{\rm FB}-\bar{A}_{\rm FB}$}
\end{overpic}
\end{center}
\caption{CP violation for the forward and backward asymmetries
of $\tau \to K \pi \nu$. The charged Higgs boson mass 
is changed as $M_H({\rm GeV})=200$ (thick solid line),$250$
 (dashed line), $500$ (short dashed line)
and $700$ (solid line). The other parameters are $\tan \beta=50$,
$|r_{2 \tau \tau}|=1$ and $\theta_{2 \tau \tau}=\frac{\pi}{2}$.}
\label{CPA}
\end{figure}

The CP violation of the forward and backward asymmetry is shown 
for $K^{\mp} \pi^0$ case in Fig.~\ref{CPA}.
By taking $|r_{2 \tau \tau}|=1$, 
one can see that CP asymmetries can be as large as a few \%.
At low invariant mass region $\sqrt{s} < 900$(MeV),
the direct CP violation is negative while at 
high invariant mass region $\sqrt{s} > 900$ (MeV)
, the CP violation is positive.
The sign is correlated to $\sin \delta_{st}$ as can be seen
in Eq.~(\ref{FBACPE}). From Fig.~\ref{phase}, we can see
$\sin \delta_{st}$ also changes its sign around $\sqrt{s}=900$(MeV).
We also change the charged Higgs boson mass. For $M_H > 500$ (GeV),
CP violation is suppressed to less than 1 $\%$.
In Fig.~\ref{CPALL}, we also show the CP asymmetries for
$K \eta$ and $K \eta'$ cases. We note the sign of the
CP violation is opposite to the sign
at high invariant mass region of $K \pi$ case. 

\begin{figure}[htbp]
\begin{center}
\begin{overpic}[width=9cm]{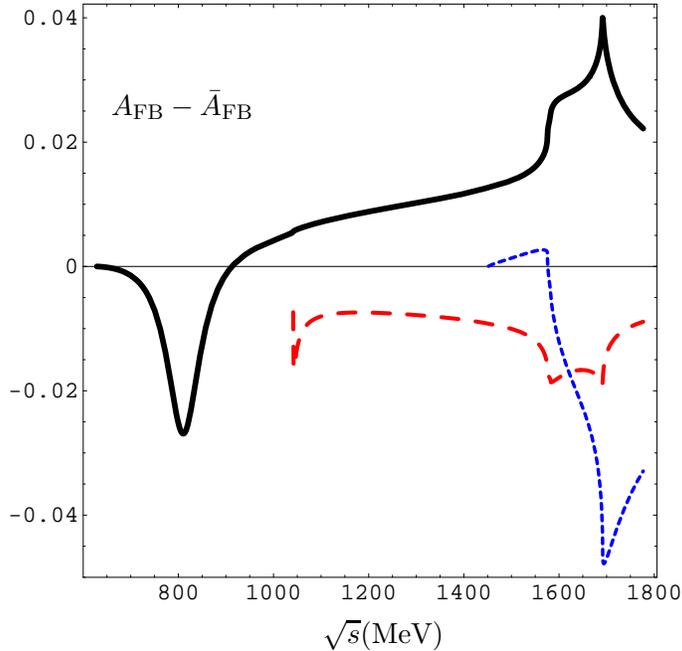}
\put(120,0){$\sqrt{s}$(MeV)}
\put(40,200){$A_{\rm FB}-\bar{A}_{\rm FB}$}
\end{overpic}
\label{CPALL}
\end{center}
\caption{CP violation for the forward and backward asymmetries
of $\tau \to K \pi \nu$ (thick solid line), $\tau \to K \eta \nu$ (dashed line) and
$ \tau \to K \eta' \nu$ (short dashed line).
 We choose the parameters as  $M_H=300$, $\tan \beta=50$
$|r_{2 \tau \tau}|=1$, $\theta_{2 \tau \tau}=\frac{\pi}{2}$ }
\label{CPALL}
\end{figure} 
\section{Conclusion and Discussion}
We have studied CP violation of 
$\tau \to K P \nu$ ($P=\pi^0, \eta,\eta'$)
decays and 
$\tau \to K_s \pi \nu$. CP violation for the forward and backward asymmetries
is computed using the two Higgs doublet model which has
the non-minimal Yukawa couplings to the charged leptons. 
The effect of the CP violation is numerically studied.
The direct CP violation depends on the strong phase shifts
$\sin(\delta_{V}-\delta_{s})$. To evaluate the phase shifts,
we use the chiral lagrangian
including the vector and the scalar resonances and take into
account of the one loop corrections.
We compute both real part and imaginary part of their
self-energies.
The divergences of the real part is subtracted properly. 
We have determined the finite renormalization constants
so that the branching fractions for
$K \pi $ and $K \eta$ modes are reproduced. 
We also take into account of 
U(1)$_A$ breaking so that $\eta_0$ and $\eta_8$ mixing 
can be incorporated.  With those improved treatment,
we can predict the hadronic invariant mass spectrum and
CP violation even at high invariant mass region for $K \pi$
case as well as the same observables of $K \eta$ and $K \eta'$
cases. 

The effect on the direct CP violation
of the non-minimal coupling is studied. For 
the non-minimal Yukawa coupling as large as that of
the standard model Yukawa coupling of $\tau$ lepton, 
we have predicted a few \% CP asymmetries within the parameter
region with $M_H = 200 \sim 300$(GeV) and with
$\tan \beta \sim 50$. 
\acknowledgments
We would like to thank 
 K. Hayasaka, K. Inami, 
T. Onogi and T. Oshima for discussion and encouragement.

This work was supported by the Korea Research Foundation Grant funded by the Korean Government (MOEHRD, Basic Research Promotion Fund, KRF-2007-C00145) and the BK21 program of Ministry of Education (K.Y.L.).

The work of T. M. was supported by KAKENHI, Grant in Aid for
Scientific Research on Priority Areas "New Development
for Flavor Physics",
No.19034008 and No.20039008, MEXT,Japan.  We also thank YITP for
support from international molecule type visitor program
by the Yukawa International
Programs for Quark-Hadron Sciences.
The preliminary results of the paper were presented in BNM2006,
YITP workshop; Towards the precise
predictions of CP violation (YITP-W-07-21),
BNM2008, New Development of
flavor physics 2008, YongPyong 2008 and HQL2008.
\appendix
\section{Scalar and vector meson nonets}
\def\sq{\sqrt{2}}
Here we show the scalar and vector meson nonets.
\bea
S=\left(\begin{array}{ccc} 
  \frac{a^0}{2}+\frac{\sigma}{2}
 & \frac{a^+}{\sq}& \frac{\kappa^+}{\sq} \\
 \frac{a^-}{\sq} & -\frac{a^0}{2} +
\frac{\sigma}{2} & \frac{\kappa^0}{\sq} \\
\frac{\kappa^-}{\sq} & \frac{\overline{\kappa^0}}{\sq} & 
f^0 \end{array}
\right), \quad 
V=\left(\begin{array}{ccc} 
  \frac{\rho}{2}+\frac{\omega}{2}
 & \frac{\rho^+}{\sq}& \frac{K^{\ast +}}{\sq} \\
 \frac{\rho^-}{\sq} & -\frac{\rho^0}{2} +
\frac{\omega}{2} & \frac{K^{\ast 0}}{\sq} \\
\frac{K^{\ast -}}{\sq} & \frac{\overline{K^{\ast 0}}}{\sq} & 
\phi \end{array}
\right).
\eea
\section{Functions $R^{(n)}_{PQ}$}
The function $R^{(n)}_{PQ}$ is defined as,
\bea
R^{(n)}_{PQ}&=&\int_{0}^{1} dx x^n 
\log(x^2-(1+\frac{\Delta_{PQ}}{s})x+\frac{M_P^2}{s}-i \epsilon).
\eea
There are two regions for $s$ of interests depending on 
below the threshold, i.e., (1)
$(M_P-M_Q)^2 \le s \le (M_P+M_Q)^2$ or the above threshold (2)
$(M_P+M_Q)^2 \ge s$.\\
For the case (2), 
\bea
R^{(n)}_{PQ}= \int_0^1  x^n \log((x-a)(x-b)-i \epsilon) dx,
\eea
with $a$ and $b$  are given as,
\bea
a&=& \frac{1+\frac{\Delta_{PQ}}{s}}{2}-\frac{\nu_{PQ}}{2s}, \nn \\
b&=& \frac{1+\frac{\Delta_{PQ}}{s}}{2}+\frac{\nu_{PQ}}{2s}, 
\eea
with $0 \le a,b \le 1$. We show the real part of $R_{PQ}^{(n)}$,
\bea
{\rm Re}.(R_{PQ}^{(0)})&=&-2+
\log(1-a)(1-b)-a \log\frac{1-a}{a}-b \log\frac{1-b}{b}, \nn \\
{\rm Re}.(R_{PQ}^{(1)})&=&\frac{a^2}{2} \log \frac{a}{1-a}+\frac{1}{2}
\log(1-a)-\frac{1}{4}-\frac{a}{2}+
\frac{b^2}{2} \log \frac{b}{1-b}+\frac{1}{2}
\log(1-b)-\frac{1}{4}-\frac{b}{2}, \nn \\
{\rm Re}.(R_{PQ}^{(2)})&=&-\frac{6a^2+3 a +2}{18}+\frac{1}{3} \log(1-a)
+\frac{a^3}{3} \log\frac{a}{1-a}, \nn \\
&-&\frac{6b^2+3 b +2}{18}+\frac{1}{3} \log(1-b)
+\frac{b^3}{3} \log\frac{b}{1-b}.
\eea
For the case (1), 
\bea
R^{(n)}_{PQ}&=&\int_0^1 dx x^n \log((x-\beta)^2+\alpha^2) \nn \\
&=& \int_{-\beta}^{1-\beta} dy 
(\beta+y)^n \log(y^2 +\alpha^2),
\eea
where, 
\bea
\alpha&=&\frac{\sqrt{((M_P+M_Q)^2-s)((s-(M_P-M_Q)^2)}}{2s}=
\frac{\sqrt{-\nu_{PQ}^2}}{2s}, \nn \\
\beta&=&\frac{s+\Delta_{PQ}}{2s}. 
\eea
We define the indefinite integrals,
\bea
r^{(n)}_{PQ}(y)&=& 
\int^{y} dy y^n \log(y^2 +\alpha^2).
\eea
Using the integrals, one can write,
\bea
R^{(0)}_{PQ}&=& r^{(0)}_{PQ}(1-\beta)-r^{(0)}_{PQ}(-\beta),\nn \\
R^{(1)}_{PQ}&=& r^{(1)}_{PQ}(1-\beta)-r^{(1)}_{PQ}(-\beta)
+ \beta R^{(0)}_{PQ}, \nn \\
R^{(2)}_{PQ}&=& r^{(2)}_{PQ}(1-\beta)-r^{(2)}_{PQ}(-\beta)
+2 \beta R^{(1)}_{PQ}-\beta^2 R^{(0)}_{PQ}. 
\eea
The indefinite integrals are given by, 
\bea
r^{(2)}_{PQ}&=& \frac{1}{3} (y^3 \log(y^2+\alpha^2)-\frac{2 y^3}{3}
+2 \alpha^2 y -2 \alpha^3 \arctan \frac{y}{\alpha}), \nn \\
r^{(1)}_{PQ}&=& \frac{1}{2}\left(
(y^2+\alpha^2) \log(y^2+\alpha^2)-y^2 \right), \nn \\
r^{(0)}_{PQ}&=&y \log(y^2+\alpha^2)-2 y + 2 \alpha 
\arctan \frac{y}{\alpha}.
\eea
\def\apj#1#2#3{Astrophys.\ J.\ {\bf #1}, #2 (#3)}
\def\ijmp#1#2#3{Int.\ J.\ Mod.\ Phys.\ {\bf #1}, #2 (#3)}
\def\mpl#1#2#3{Mod.\ Phys.\ Lett.\ {\bf A#1}, #2 (#3)}
\def\nat#1#2#3{Nature\ {\bf #1}, #2 (#3)}
\def\npb#1#2#3{Nucl.\ Phys.\ {\bf B#1}, #2 (#3)}
\def\plb#1#2#3{Phys.\ Lett.\ {\bf B#1}, #2 (#3)}
\def\prd#1#2#3{Phys.\ Rev.\ {\bf D#1}, #2 (#3)}
\def\ptp#1#2#3{Prog.\ Theor.\ Phys.\ {\bf #1}, #2 (#3)}
\def\pr#1#2#3{Phys.\ Rev.\ {\bf #1}, #2 (#3)}
\def\prl#1#2#3{Phys.\ Rev.\ Lett.\ {\bf #1}, #2 (#3)}
\def\prp#1#2#3{Phys.\ Rep.\ {\bf #1}, #2 (#3)}
\def\sjnp#1#2#3{Sov.\ J.\ Nucl.\ Phys.\ {\bf #1}, #2 (#3)}
\def\zp#1#2#3{Z.\ Phys.\ {\bf #1}, #2 (#3)}
\def\jhep#1#2#3{JHEP\ {\bf #1}, #2 (#3)}
\def\epjc#1#2#3{Eur. Phys. J.\ {\bf C#1}, #2 (#3)}
\def\rmp#1#2#3{Rev. Mod. Phys.\ {\bf #1}, #2 (#3)}
\def\prgth#1#2#3{Prog. Theor. Phys.\ {\bf #1}, #2 (#3)}

\end{document}